\documentclass[12pt]{article}

\usepackage{epsf}

\newcommand{\bmat}{\left(\begin{array}}
\newcommand{\emat}{\end{array}\right)}
\def\NPB#1#2#3{Nucl. Phys. B{#1} (19#2) #3}
\def\PLB#1#2#3{Phys. Lett. B{#1} (19#2) #3}

\def\PRD#1#2#3{Phys. Rev. D{#1} (19#2) #3}
\def\PRL#1#2#3{Phys. Rev. Lett. {#1} (19#2) #3}

\def\yzero{\smash{\hbox{$y\kern-4pt\raise1pt\hbox{${}^\circ$}$}}}

\def\beq{\begin{equation}}
\def\eeq{\end{equation}}
\def\beqa{\begin{eqnarray}}
\def\eeqa{\end{eqnarray}}

\def\-{\hphantom{-}}

\def\s2{\frac{1}{\sqrt2}}

\def\beq{\begin{equation}}
\def\eeq{\end{equation}}
\def\beqa{\begin{eqnarray}}
\def\eeqa{\end{eqnarray}}

\def\IF{\relax{\rm I\kern-.18em F}}
\def\II{\relax{\rm I\kern-.18em I}}
\def\IT{\relax{\rm I\kern-.18em T}}
\def\IP{\relax{\rm I\kern-.18em P}}
\def\IC{\relax\hbox{\kern.25em$\inbar\kern-.3em{\rm C}$}}
\def\IR{\relax{\rm I\kern-.18em R}}

\def\cn{{\cal N}}

\def\Dsl{\,\raise.15ex\hbox{/}\mkern-13.5mu D} 
\def\IZ{Z\kern-.4em  Z}

\catcode`\@=11
\newdimen\@rotdimen
\newbox\@rotbox

\def\@vspec#1{\special{ps:#1}}
\def\@rotstart#1{\@vspec{gsave currentpoint currentpoint translate
   #1 neg exch neg exch translate}}
\def\@rotfinish{\@vspec{currentpoint grestore moveto}}
\def\@rotr#1{\@rotdimen=\ht#1\advance\@rotdimen by\dp#1%
   \hbox to\@rotdimen{\hskip\ht#1\vbox to\wd#1{\@rotstart{90 rotate}%
   \box#1\vss}\hss}\@rotfinish}
\def\@rotl#1{\@rotdimen=\ht#1\advance\@rotdimen by\dp#1%
   \hbox to\@rotdimen{\vbox to\wd#1{\vskip\wd#1\@rotstart{270 rotate}%
   \box#1\vss}\hss}\@rotfinish}%
\def\@rotu#1{\@rotdimen=\ht#1\advance\@rotdimen by\dp#1%
   \hbox to\wd#1{\hskip\wd#1\vbox to\@rotdimen{\vskip\@rotdimen
   \@rotstart{-1 dup scale}\box#1\vss}\hss}\@rotfinish}%
\def\@rotf#1{\hbox to\wd#1{\hskip\wd#1\@rotstart{-1 1 scale}%
   \box#1\hss}\@rotfinish}%
\def\rotate{\@ifnextchar[{\@rotate}{\@rotate[l]}}
\def\@rotate[#1]#2{\setbox\@rotbox=\hbox{#2}\@nameuse{@rot#1}\@rotbox}

\catcode`\@=12

\topmargin
-1.5cm
\textwidth
15.5cm
\textheight
23.5cm
\oddsidemargin
0.7cm
\evensidemargin
1.2cm

\begin{document}

\makeatletter
\@addtoreset{equation}{section}
\makeatother
\renewcommand{\theequation}{\thesection.\arabic{equation}}
\pagestyle{empty}
\rightline{FTUAM-01/09; IFT-UAM/CSIC-01-15}
\rightline{\tt hep-th/0105155}
\vspace{0.5cm}
\begin{center}
\LARGE{Getting  just the Standard Model at Intersecting Branes  \\[10mm]}
\large{
L.~E.~Ib\'a\~nez, F. Marchesano and R. Rabad\'an 
\\[2mm]}
\small{
 Departamento de F\'{\i}sica Te\'orica C-XI
and Instituto de F\'{\i}sica Te\'orica  C-XVI,\\[-0.3em]
Universidad Aut\'onoma de Madrid,
Cantoblanco, 28049 Madrid, Spain.
\\[9mm]}
\small{\bf Abstract} \\[7mm]
\end{center}

\begin{center}
\begin{minipage}[h]{14.0cm}

We present what we believe are the first specific string 
(D-brane) constructions
whose low-energy limit yields just a three generation
$SU(3)\times SU(2)\times U(1)$
standard model with no extra fermions nor $U(1)$'s   (without any further
effective  field theory assumption). 
In these constructions the number of generations is given by the 
number of colours.
The Baryon, Lepton and Peccei-Quinn 
symmetries are necessarily gauged and their anomalies cancelled
by a generalized Green-Schwarz mechanism.
 The corresponding 
gauge bosons become massive but
 their presence guarantees automatically
proton stability.
There are necessarily three right-handed neutrinos and
neutrino masses can only be of Dirac type. They are
naturally small as a consequence of a PQ-like symmetry.
There is a Higgs sector which is somewhat  similar to that
of the MSSM and the scalar potential parameters have a geometric
interpretation in terms of brane distances and intersection
angles.  Some other physical implications of these constructions 
are discussed.

\end{minipage}
\end{center}
\newpage
\setcounter{page}{1}
\pagestyle{plain}
\renewcommand{\thefootnote}{\arabic{footnote}}
\setcounter{footnote}{0}


\section{Introduction}

If string theory is to describe the observed physics, it should 
be possible to find string configurations containing the 
observed standard model (SM). In the last fifteen years it
has been possible to construct string vacua with massless 
sector close to the SM with three quark/lepton generations \cite{phen}.
 However all string constructions 
up to now lead to extra massless fermions and/or gauge bosons
in the low energy spectrum. The hidden reason for this fact 
is that all those constructions contain extra $U(1)$ 
(or non-Abelian) gauge symmetries beyond 
$SU(3)\times SU(2)\times U(1)_Y$ and the cancellation of gauge anomalies 
requires in general the presence of extra chiral fermions beyond
the spectrum of the SM. The usual procedure in the
literature is then 
to  abandon the string theory techniques and use instead the 
low-energy effective Lagrangian below the string scale. Then 
one tries to find some scalar field direction in which 
all extra gauge symmetries are broken and extra fermions 
become massive. This requires a very complicated 
model dependent   analysis
of the structure of the scalar potential and Yukawa couplings
and, usually, the necessity of unjustified simplifying 
assumptions.  For example, there is a lot of arbitrariness in the
choice of scalar flat directions and the physics varies drastically
from one choice to another. Fundamental properties like 
proton stability typically result from the particular choice
of scalar flat direction.

It is clear that it would be nice to have some string constructions 
with massless spectrum identical to that of the SM and 
with gauge group just $SU(3)\times SU(2)\times U(1)_Y$ 
already at the string theory level, without any effective field theory
elaboration. This could constitute an important first step to
the string theory description of the observed world. But it could 
also give us some model independent understanding of some of the 
mysteries of the SM like generation-replication or 
the stability of the proton.

In the present article we report on the first such string constructions
yielding just the SM massless spectrum from the start.
We consider the SM gauge group as arising from four sets of D-branes
wrapping cycles on compactified (orientifolded) Type II string theory. 
At the intersections of the branes live chiral fermions to be
identified with quarks and leptons. In order to obtain just the 
observed three generations of quarks and leptons the D-brane
cycles have to intersect the appropriate number of times 
as shown in eq.(\ref{intersec2}). 
Models with quarks and leptons living at D-brane intersections 
 were already considered in refs.\cite{bgkl,afiru,afiru2,bkl}.
In those papers three generation models were obtained but
involving either extra chiral fermions and $U(1)$'s beyond the SM
\cite{afiru,afiru2} or else an extended gauge group beyond the 
SM \cite{bkl}. As we said, the models we report here  
have just the SM gauge group  and there are no extra chiral fermions 
nor $U(1)$ gauge bosons. 

The particular examples we discuss consist on sets of 
D6-branes wrapping on an orientifolded six-torus \cite{bachas,bgkl}
in the presence of a background NS B-field \cite{bkl,bfield,flux}.
We classify the D6-brane cycles yielding the SM spectrum.
We find certain families of models which depend on a few integer 
parameters. The analysis of the $U(1)$ gauge anomalies in the 
constructions, along the lines discussed in ref.\cite{afiru}, 
is crucial in obtaining the correct SM structure. In particular, 
we find that the four original gauge $U(1)$ symmetries 
can be identified with  Baryon number, Lepton number, Peccei-Quinn symmetry 
and hypercharge (or linear combinations thereof). There are
 Wess-Zumino-like $B\wedge F$ couplings involving RR fields and the Abelian 
gauge bosons. Due to these couplings, three of the $U(1)$'s become
massive (of order the string scale) and only a $U(1)$ remains 
in the massless spectrum. We discuss the conditions under which
the remaining $U(1)$ symmetry is the standard hypercharge.
This substantially constraints the structure of the configurations
yielding $SU(3)\times SU(2)\times U(1)_Y$ as the only gauge group.

The D6-brane configurations we study are non-supersymmetric. 
However we show explicitly, extending a previous analysis in 
ref.\cite{afiru},  that for wide
ranges of the  parameters (compactification radii) there are no tachyonic
scalars at any of the intersections. Thus at this level the configurations
are stable. We also show that for certain values of the
geometrical data (some brane distances and intersection angles)
the required Higgs scalar multiplets may appear in the light 
spectrum. The obtained Higgs sector is quite analogous to the 
one appearing in the minimal supersymmetric standard model (MSSM).
But in our case the parameters of the scalar potential have 
a geometrical interpretation in terms of the brane distances and angles.
Since the models are non-supersymmetric, the string scale should then be of
order 1-few TeV.

We find that all the SM configurations obtained have a number of common
features  which seem to be quite model independent and could be
a general property of any string model yielding 
{\it just} the SM spectrum:

\begin{itemize}

\item
A first property is the connection between the number of generations
and the number of colours. Indeed, in order to cancel anomalies while having
complete quark/lepton
generations,  the number of generations in the provided 
constructions must be equal to the number 
of colours, three. This is quite an elegant explanation for the
multiplicity of generations: there is no way in which we could 
construct e.g., a D-brane model with just one generation.

\item
Baryon number is an exact symmetry in perturbation theory.
Indeed, we mentioned that Baryon number is a gauged $U(1)$ symmetry
of the models. Although naively anomalous, the anomalies are cancelled 
by a generalized Green-Schwarz 
\cite{gs} mechanism which at the same time gives 
a mass to the corresponding gauge boson. The corresponding $U(1)$ symmetry
remains as an effective global symmetry in the effective Lagrangian.
This is a remarkable simple explanation for the observed stability of the
proton. Indeed, the standard explanation for the surprisingly
high level of stability of the proton is to suppose that the 
scale of baryon number violation is extremely large, larger than 
$10^{16}$ GeV. This requires to postpone the scale of a more
fundamental theory to a scale at least as large as $10^{16}$ 
GeV. In our present context the scale of fundamental physics
may be as low as 1 TeV without any problem with proton stability.
Thus, in particular, this provides a natural explanation 
for proton stability in brane-world models with a low (of order 1-few TeV)
string scale \cite{aadd,otherbw}.

\item 
Lepton number is an exact symmetry in perturbation theory. 
Again, Lepton number is a gauged symmetry and remains as a
global symmetry in the effective action. This has the 
important consequence that Majorana neutrino masses are
forbidden. On the other hand another generic feature 
of our class of models is the necessary presence of 
three generations of right-handed neutrinos (singlets under
hypercharge). In general Dirac neutrino masses may be present
and neutrinos may oscillate in the standard way since it
is only the diagonal lepton number $L=L_e+L_{\mu }+L_{\tau }$ which is 
an exact symmetry. 

\item
There is a generation-dependent  Peccei-Quinn symmetry \cite{pq} which is
also gauged and thus in principle remains as a global $U(1)$ symmetry in the
effective Lagrangian.

\end{itemize}

In addition to these general properties, our class of models have
other interesting features.
We already mentioned that Higgs fields appear under certain 
conditions and that the Higgs sector comes in sets analogous to that of
the  Minimal Supersymmetric Standard Model (MSSM).
 There is no gauge coupling unification, 
the size of each gauge coupling constant
is inversely proportional to
the volume wrapped by the corresponding brane. Thus it seems one 
can reproduce the observed size of the gauge couplings by 
appropriately varying the compact volume.
Yukawa
couplings may be computed in terms of the area of the world-sheet
stretching among the different branes. The dependence on this
area is exponential, which may give an understanding \cite{afiru2}  of   
the hierarchy of fermion masses.  
In the case of neutrino masses, one finds that they may be naturally small 
as a consequence of the PQ-like symmetry.

The structure of the present paper is as follows.
In chapter 2 we describe the  general philosophy in order 
to obtain the minimal spectrum of the SM at the intersections
of wrapping D-branes. We also discuss the connection between
the number of colours and generations in these constructions
as well as the general structure of $U(1)$ anomaly cancellation.
In chapter 3  we review  the case of intersecting D6-branes wrapping on 
a six-torus and describe the spectrum as well as the general
Green-Schwarz mechanism in these theories. 
The search for specific D6-brane configurations yielding
just the SM spectrum is carried out in chapter 4. We present 
a classification of such type of models and study tadpole
cancellation conditions. We also find the general conditions
under which the $U(1)$ remaining light is the standard hypercharge.

In chapter 5 we study the stability of the brane configurations.
Specifically we show the absence of tachyons for wide ranges of
the compact radii. The spectrum of massive particles  
from KK and winding states is discussed in chapter 6.
 The appearance of Higgs fields in the light spectrum is
discussed in chapter 7. We discuss the multiplicity of those 
scalars as well as the general structure of the mass terms
appearing in the scalar potential. A brief discussion of 
the gauge coupling constants and the Yukawa couplings is offered in 
chapter 8. We leave chapter 9 for some final  comments
and conclusions.

\section{The standard model intersection numbers}

In our search for a string-theory description of the 
standard model (SM) we are going to consider configurations
of D-branes wrapping on cycles on the six extra dimensions, which we will
 assume to be compact. Our aim is to find 
configurations with just the SM group $SU(3)\times SU(2)\times U(1)_Y$
as a gauge symmetry and with three generations of fermions 
transforming like the five representations:
\beqa
& Q_L^i=(3,2,1/6)\ ; \ U_R^i =({\bar 3},1,-2/3)\ ;\
D_R^i=({\bar 3},1,1/3)\ ; \nonumber
\\  & L^i=(1,2,-1/2)\ ;\ E_R^i=(1,1,1)  \ .
\label{gensm}
\eeqa
Now, in general, D-branes will give rise to $U(N)$ gauge factors in their
world-volume, rather than $SU(N)$. Thus, if we have $r$  different stacks
with $N_i$ parallel branes  we will expect gauge groups in general of the form 
$U(N_1)\times U(N_2)\times .... \times U(N_r)$. At points where the
D-brane cycles intersect one will have in general massless 
fermion fields transforming like bifundamental representations,
i.e., like $(N_i,{\bar N}_j)$ or $(N_i,N_j)$. Thus the  idea will be
to identify these fields with the SM fermion fields.

A first obvious idea is to consider
three types of branes giving rise in their world-volume to
a gauge group $U(3)\times U(2)\times U(1)$. This in general turns out not to 
be sufficient. Indeed, as we said,  chiral fermions like those in the SM 
 appear from open strings stretched between D-branes
with intersecting cycles. 
 Thus e.g., the left-handed quarks $Q_L^i$
can only appear from open strings stretched between the $U(3)$ branes
and the $U(2)$ branes. In order to get the right-handed
leptons $E_R^i$ we would need a fourth set with one brane giving rise
to an additional $U(1)'$: the right-handed leptons would come from
open strings stretched between the two $U(1)$ branes. Thus we will be
forced to have a minimum of four stacks of branes with
$N_1=3,N_2=2,N_3=1$ and $N_4=1$ yielding a
$U(3)\times U(2)\times U(1)\times U(1)$ gauge group
\footnote{Although apparently such a structure would yield four gauged 
$U(1)$'s, we show below that we expect three of these $U(1)$'s to
become massive and decouple from the low-energy spectrum}.

In the class of models we are considering the fermions 
come in bifundamental representations:
\beq
\sum_{a,b}
n_{ab}(N_a,{\overline N}_b)+m_{ab}  (N_a,N_b) + n_{ab}^* ({\overline N}_a,
N_b)
+m_{ab}^*
({\overline N}_a , {\overline N}_b)  \ .
\label{bifundamentals}
\eeq
where here $n_{ab},n_{ab}^*,m_{ab},m_{ab}^*$ are integer non-negative
coefficients which are model dependent
\footnote{ In some orientifold cases there
may appear fermions transforming like antisymmetric or symmetric
representations. For the case of the SM group those states would 
give rise to exotic chiral fermions which have not been  observed.
Thus we will not consider these more general possibilities any further.}.
In all D-brane models strong constraints appear from
 cancellation of Ramond-Ramond (RR) tadpoles. 
 Cancellation of tadpoles  
 also guarantees the cancellation of gauge 
anomalies. In the case of the D-brane models here discussed
anomaly cancellation just requires that there should be
as many $N_a$ as ${\overline N}_a$ representations for any $U(N_a)$
group.

An important fact for our discussion later is that tadpole
cancellation conditions impose this constraint even if the  
gauge group is $U(1)$ or $U(2)$. 
The constraint in this case turns out to be  required
for the cancellation of $U(1)$ anomalies.
Let us now apply this to a possible
D-brane model yielding  
$U(3)\times U(2)\times U(1)\times U(1)$ gauge group.
Since  $U(2)$ anomalies have to cancel we will make a distinction
between $U(2)$ doublets and antidoublets. Now, in the SM only
left-handed quarks and leptons are $SU(2)$ doublets. Let us assume
to begin with that the three left-handed quarks $Q_L^i$ were
antidoublets $(3,{\bar 2})$. Then there are altogether 
9 anti-doublets and $U(2)$ anomalies would
never cancel with just three generations of left-handed leptons.
Thus all models in which all left-handed quarks are $U(2)$ doublets
(or antidoublets) will necessarily require the presence in the spectrum of 
9 $U(2)$ lepton doublets (anti-doublets)
\footnote{Indeed this can be checked for example in the semi-realistic models of
ref.\cite{evenmore,aiqu,afiru,afiru2,bjl}.}. 
There is however a simple  way to 
cancel $U(2)$ anomalies sticking to the fermion content of the
SM. They cancel if  two of the left-handed quarks 
are antidoublets and the third one is a doublet
\footnote{In ref.\cite{bkl} a $U(3)\times U(2)_L\times U(2)_R
\times U(1)$ model  of these characteristics was built.}. Then
there is a total of  six doublets and antidoublets 
and $U(2)$ anomalies will cancel.

 Notice that in this case 
it is crucial that the {\it  number of generations equals the number
of colours}. There is no way to build a D-brane configuration 
with the gauge group of the SM and e.g., just one 
complete quark/lepton generation. Anomalies (RR tadpoles) cannot possibly
cancel\footnote{ It is however in principle possible to get 2-generation
models simply by assuming that one generation has only $U(2)$ doublets
and the other antidoublets. On the other hand we have made an analysis
like that in chapter 4 for the case of two generations and have found that in
D6-brane toroidal models it is not possible to have just the SM group, there
is always an additional  $U(1)$ beyond hypercharge which is present 
in the massless spectrum. This is also related to the fact that in the
2-generation case there is only one anomalous $U(1)$ which is $B+L$.
 Thus, at least within that class of models, the 
minimal configuration with {\it just} the SM group requires 
at least three generations.}. 
  We find
this connection between the number of 
generations and colours quite attractive.

From the above discussion we see that, if we want to stick to
the particle content of the minimal SM , we will need to 
consider string configurations in which {\it both}
types of bifundamental fermion representations, 
$(N_a, {\bar N}_b)$ and $(N_a, N_b)$ appear in the massless 
spectrum.  This possibility is familiar from Type II 
orientifold \cite{orientold,orientnew}
 models in which the world-sheet of the 
string is modded by some operation of the form
$\Omega {\it R}$, where $\Omega $ is the world-sheet
parity operation and  
${\it R}$ is some geometrical action. Bifundamental representations
of type $(N_a, {\bar N}_b)$ appear from open strings stretched 
between branes $a$ and $b$ whereas those of type 
$(N_a, N_b)$ appear from those going between branes $a$ and  $b^*$,
the latter being the mirror of the $b$ brane under the $\Omega {\it R}$
operation. We will thus from now on assume that we are considering 
string and brane  configurations  in which both types
of bifundamentals appear. Specific examples will be considered in the 
following sections.

It is now clear what are we looking for. We are searching for 
brane configurations with four stacks of branes yielding 
an initial  $U(3)\times U(2)\times U(1)\times U(1)$
gauge group. They wrap cycles $\Pi_i$, $i=a,b,c,d$ and intersect with each
other a number of times given by the intersection numbers
$I_{ij}=\Pi_i . \Pi_j $. In order to reproduce the desired
fermion spectrum (depicted in table 1) the intersection numbers should be
\footnote{An alternative with $I_{ab}=2, I_{ab*}=1, 
I_{bd}=3, I_{bd*}=0$ gives equivalent spectrum.}:
\beqa
I_{ab}\ & = &  \ 1 \ ;\ I_{ab*}\ =\ 2  \nonumber \\
I_{ac}\ & = &  \ -3 \ ;\ I_{ac*}\ =\ -3  \nonumber \\
I_{bd}\ & = &  \ 0  \ ;\ I_{bd*}\ =\ -3  \nonumber \\
I_{cd}\ & = &  \ -3 \ ;\ I_{cd*}\ =\ 3
\label{intersec2}
\eeqa
all other intersections vanishing. 
Here a negative number denotes that the
corresponding fermions should have opposite chirality to those
with positive intersection number.
As we discussed, cancellation of $U(N_i)$ anomalies requires:
\beqa
\sum_j\, I_{ij}\, N_j = 0
\label{anomdsix}
\eeqa
which is indeed obeyed by the spectrum of table 1, although
to achieve this cancellation we have to 
 add three fermion singlets $N_R$. As shown below these have
quantum numbers of right-handed neutrinos (singlets under hypercharge).
Thus this is a first prediction of the present approach: 
{\it right-handed neutrinos must exist}.

\begin{table}[htb] \footnotesize
\renewcommand{\arraystretch}{1.25}
\begin{center}
\begin{tabular}{|c|c|c|c|c|c|c|c|}
\hline Intersection &
 Matter fields  &   &  $Q_a$  & $Q_b $ & $Q_c $ & $Q_d$  & Y \\
\hline\hline (ab) & $Q_L$ &  $(3,2)$ & 1  & -1 & 0 & 0 & 1/6 \\
\hline (ab*) & $q_L$   &  $2( 3,2)$ &  1  & 1  & 0  & 0  & 1/6 \\
\hline (ac) & $U_R$   &  $3( {\bar 3},1)$ &  -1  & 0  & 1  & 0 & -2/3 \\
\hline (ac*) & $D_R$   &  $3( {\bar 3},1)$ &  -1  & 0  & -1  & 0 & 1/3 \\
\hline (bd*) & $ L$    &  $3(1,2)$ &  0   & -1   & 0  & -1 & -1/2  \\
\hline (cd) & $E_R$   &  $3(1,1)$ &  0  & 0  & -1  & 1  & 1   \\
\hline (cd*) & $N_R$   &  $3(1,1)$ &  0  & 0  & 1  & 1  & 0 \\
\hline \end{tabular}
\end{center} \caption{ Standard model spectrum and $U(1)$ charges 
\label{tabpssm} }
\end{table}

The structure of the $U(1)$ gauge fields is very important in what follows. 
The following important points are in order:

{\bf 1)}
The four $U(1)$ symmetries $Q_a$, $Q_b$, $Q_c$ and $Q_d$ have clear
interpretations in terms of known global symmetries of the standard model. 
Indeed $Q_a$ is $3B$, $B$ being the baryon number and $Q_d$ is nothing
but (minus)lepton number. Concerning $Q_c$, it is twice $I_R$, the third
component of right-handed weak isospin familiar from left-right
symmetric models. Finally $Q_b$ has the properties of a Peccei-Quinn
symmetry (it has mixed $SU(3)$ anomalies).
We thus learn that {\it all these known global symmetries of the SM 
are in fact gauge symmetries} in this class of theories.

{\bf 2)}
The mixed anomalies $A_{ij}$ of these four $U(1)_i$'s with the non-Abelian
groups $SU(N_j)$
are given by:
\beq
A_{ij} \ =\ {1\over 2} \  ( I_{ij}-I_{i*j})\ N_i \ .
\label{anomix}
\eeq
It is easy to check that $(Q_a+3Q_d)$ (which is $3(B-L)$) 
and $Q_c$ are free of triangle anomalies. In fact the hypercharge
is given by the linear combination:
\beq
Q_Y \ =\ {1\over 6} Q_a\ -\ {1\over 2} Q_c \ +\ {1\over 2} Q_d
\label{hyper}
\eeq
and is, of course, anomaly free. On the other hand the other
orthogonal combinations $(3Q_a-Q_d)$ and $Q_b$ have triangle anomalies.
Of course, if the theory is consistent these anomalies should somehow
cancel. What happens is already familiar from heterotic compactifications
\cite{dsw} and  Type I and Type II theories in six \cite{sagnan} and four
\cite{iru} dimensions. 
There will be closed string modes
coupling to the gauge fields giving rise to  a generalized Green-Schwarz
mechanism. This will work in general as follows.
Typically there are RR two-form  fields  $B_{\alpha}$ with couplings 
to  the $U(1)_i$ field strengths:
\beq 
\sum_{\alpha} c_{\alpha}^i  \ B_{\alpha} \wedge  tr (F^i)
\label{gsuno}
\eeq
and in addition 
there are  couplings of the Poincare  dual scalars $\eta_{\alpha}$ 
of the $B_{\alpha}$ fields:
\beq
\sum_{\alpha} d_{\alpha}^j \eta_{\alpha} tr(F^j\wedge F^j)
\label{gsdos}
\eeq
 where $F^j$ are the field strengths of any of the gauge groups.
The combination of both couplings cancels the 
mixed $U(1)_i$ anomalies with any other group $G_j$ 
as:
\beq
A_{ij}\ +\ \sum_{\alpha} c_{\alpha}^i d_{\alpha}^j \ =\ 0   \ .
\label{gstres}
\eeq
Notice two important points:

a) Given  $i,j$, for anomalies to cancel both $c_{\alpha}^i$ and 
$d_{\alpha}^j$ have to be  non-vanishing for some ${\alpha}$.

b) The couplings in (\ref{gsuno}) give masses to some 
combinations of $U(1)$'s. This always happens for 
anomalous $U(1)$'s since in this case both $c_{\alpha}^i$ and
$d_{\alpha}^j$ are necessarily  non-vanishing. However it may also happen
for some anomaly-free $U(1)$'s for which the corresponding 
combination of $\eta_{\alpha}$ fields does not couple to any $F\wedge F$
piece.

In our case  the $(3Q_a-Q_d)$ and $Q_b$ gauge bosons will become 
massive. On the other hand the other two
anomaly free combinations (including hypercharge) may be massive or not,
depending on the couplings $c_{\alpha}^i$.
Thus in order to really obtain a standard model gauge group
with the right standard hypercharge we will have to insure that
it does not couple to any closed string mode
which would render it massive, i.e., one should have
\beq
\sum _{\alpha} \ ( {1\over 6} c_{\alpha}^a \ -\
{1\over 2}c_{\alpha}^c \ +\  {1\over 2} c_{\alpha}^d \ )\  =\ 0
\label{ortohiper}
\eeq
This turns out to be an important constraint in the specific 
models constructed in the following sections. 
But an important conclusion is that in those models generically only
three of the four $U(1)$'s can become massive 
and that in a large subclass of models it is the SM hypercharge 
which remains massless. Thus even though we started with 
four $U(1)$'s we are left at the end of the day with just the 
SM gauge group.

Let us also remark that the symmetries whose gauge boson become
massive {\it  will 
persist in the low-energy spectrum as global symmetries }. 
This has important obvious consequences, as we will discuss below.

Up to now we have been relatively general and perhaps a structure like
this may be obtained in a variety of string constructions. We believe 
that the above discussion identifies in a clear way what we should
be looking for in order to get a string construction with a 
massless sector identical to the SM. In the following sections we will be more
concrete and show how this philosophy may be followed in a 
simple setting. Specifically, we will be considering Type IIA D6-branes  
wrapping at angles \cite{bgkl} 
on a six torus $T^6$. We will see how even in such
a simple setting one can obtain the desired structure.

\section{D6-branes intersecting at angles}

Let us consider D6-branes wrapping homology 3-cycles on a 
six dimensional manifold $\cal M$. Some general  features of 
this construction do not depend on the specific choice of metric 
on this space but only on the homology of these 
three cycles and its intersection 
form. More concrete problems, as the supersymmetry preserved by 
the configuration or the presence of tachyons on it, 
will depend on the metric. We will discuss first some of 
these abstract properties to proceed later to review the toroidal case 
in detail.

Two D6-branes on three-cycles will intersect generically 
at a finite number of points and those intersections will be four-dimensional.
 The intersection number depends 
on the homology class of the cycles. Deforming the D-branes 
within the same homology class the intersection number does not change. Let 
us take a basis for the $H_3({\cal M},\IZ)$,  $\Sigma_i$, 
where $i =1,\dots, b_3$ and $b_3$ is the correspondent 
Betti number.
Let us call $C_{ij}$ the intersection number of the 
cycles $\Sigma_i$ and $ \Sigma_j$.
Some properties will depend only on the 
homology of the cycles, $\Pi_a$, where the D-branes are living:

\begin{itemize}

\item there is a massless $U(1)$ field on each brane 
that can be enhanced to a $U(N)$ if N of these D-brane 
coincide. Some of the $U(1)$ factors will be massive 
due to the WZ couplings.

\item There is a chiral fermion \cite{bdl,angles} at each 
four-dimensional intersection between 
the cycles $\Pi_a$, $\Pi_b$ transforming in the bifundamental 
of $U(N_a) \times U(N_b)$, where the specific chirality depends 
on the sign of the intersection, and the number represents the multiplicity.

\item There are some conditions related to the propagation of RR 
massless closed string fields on the compact space. 
These are the RR tadpoles. These tadpoles can be expressed in a very 
simple way: the sum of the cycles where the branes are living must vanish 
\cite{probes,afiru}:
\beq
\sum_a^K N_a \Pi_a = 0 
\label{tadpole}
\eeq
In the presence of additional sources for RR charge (e.g.,
orientifold planes) one should add the corresponding 
contribution (see the toroidal example below).  
RR tadpoles imply the cancellation of all abelian gauge 
anomalies. Tadpoles of the particles in the NSNS sector are 
in general not cancelled (unless the configuration preserves 
some supersymmetry). 

\item Some of the above $U(1)$ gauge fields will be anomalous 
and the anomalies are expected to cancel in the way
described in the previous section. The triangle 
anomaly can be computed directly from the chiral spectrum, and after 
imposing tadpole cancellation conditions 
(\ref{tadpole}) one gets for  the $SU(N_b)^2 \times U(1)_a$ anomaly:
\beq
A_{ab} = \frac{1}{2} N_a (\Pi_a . \Pi_b)  
\label{U(1)}
\eeq
\end{itemize}

where $(\Pi_a . \Pi_b)$ is the intersection 
number of the $\Pi_a$ and $\Pi_b$ cycles.

As we have mentioned above other properties as the presence 
or absence of tachyons, the supersymmetries preserved 
\cite{witten,kachru,bbh} by the 
system of D-branes, the massive spectrum, etc.   will depend 
on the specific choice of metric, B-field and other background values.

\subsection{D6-branes on a torus}


Let us consider a particular example of the above 
ideas : D6-branes wrapping a three cycle on a six dimensional 
torus. A more specific choice consists on a factorization of the six 
dimensional torus into $T^2 \times T^2 \times T^2$. We can wrap a D6-brane 
on a 1-cycle of each $T^2$ so it expands a three dimensional cycle on 
the whole $T^6$ \footnote{Notice that this is not the most general 
cycle because this type of configuration only expands the 
$(H_1(T^2,\IZ))^3$ sublattice of the whole $H_3(T^6,\IZ)$.
 The construction we are considering has dimension 8 while 
the $H_3(T^6,\IZ)$ lattice has dimension $b_3 = 20$. One type of 
three cycle we are not taking into account is, for instance, 
the one that wraps the first $T^2$ and one cycle on one of the other tori.}. 
Let us denote by $(n_a^i,m_a^i)$ the wrapping numbers of the $D6_a$-brane on 
the $i$-th $T^2$. We refer the reader to  refs.\cite{bgkl,bkl} where these kind of configurations have been studied in detail.


The metric on each $T^2$ is constant and can be parametrized
by a couple of
fields: the complex structure $U$ and the 
complexified Kahler form $J$. The complexification of the 
Kahler form is done by taking in addition with the area the B-field value.  
The above models  have  a T-dual description in terms 
of D9-branes with magnetic fluxes. The 
 T-duality transformation can be carried out 
in each $T^2$ separately. A D-brane wrapped on a $(n,m)$ 
is mapped to a $U(n)$ field with a constant field strength $F$ whose 
first Chern class is m.
The D6-brane boundary  conditions
\begin{eqnarray}
\sin \vartheta_a^I \partial _{\sigma}X_1^{I}- \cos\vartheta_a^I \partial
_{\sigma}X_2^I & =& 0 \nonumber\\
\sin \vartheta _a^I \partial _{\tau}X_2^{I}-\cos\vartheta_a^I \partial
_{\tau}X_1^I &=& 0
\end{eqnarray}
are  translated into \cite{bgkl}
\begin{eqnarray}
\partial _{\sigma}X_1^{I}- F_a^I \partial
_{\tau}X_2^I & =& 0 \nonumber\\
\partial _{\sigma}X_2^{I}- F_a^I \partial
_{\tau}X_1^I &=& 0
\end{eqnarray}
in the T-dual picture \footnote{See \cite{bgkl,bkl,flux,torons,bachas}.}. 
The flux $F$ is related to the angle between 
the brane and the direction where T-duality is performed
$F_a^I = \cot{\vartheta_a^I}$.
T-duality on the three two tori takes the D6-brane system 
to a system with D9-branes with fluxes and 
the complex structure and complexified Kahler form 
interchanged. On this paper we will use the D6-branes 
at angles picture because it is easier to visualize the 
specific brane constructions.

\subsubsection{Orientifolds}

\begin{figure}
\centering
\epsfxsize=5in
\hspace*{0in}\vspace*{.2in}
\epsffile{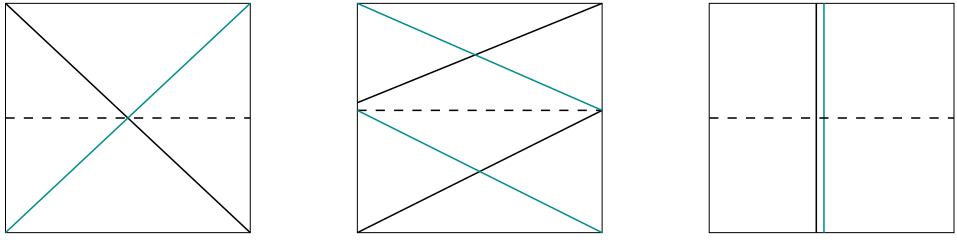}
\caption{\small The $\Omega{\cal R}$ world sheet parity 
takes one set of branes specified by $(n_i,m_i)$ to  
another set $(n_i,-m_i)$. The dashed line represent 
the direction where the O6-plane lives.} 
\label{orientifold} 
\end{figure}

Let us start from Type I string theory with branes 
with fluxes on a six dimensional torus
\cite{bachas,bgkl}. Perform a 
T-duality on the $x^4$, $x^6$, $x^8$ directions. 
The world sheet parity $\Omega$ is mapped into  
$\Omega{\cal R}$ where ${\cal R}$ is a reflection on the 
T-dualized coordinated $x^4$, $x^6$ and $x^8$. The 
D9-branes with fluxes are translated into D6-branes 
at angles. Consistency with the $\Omega{\cal R}$ 
symmetry requires the D6-branes to be in 
pairs: if $(n,m)$ are the wrapping numbers of a brane 
along a two dimensional torus, there must be a 
$\Omega{\cal R}$ partner wrapping the cycle $(n,-m)$ 
(See figure \ref{orientifold}). Let us denote by 
$\Omega{\cal R}D_a$6-brane the $\Omega{\cal R}$ image 
of the brane $D_a$6-brane. 

The spectrum can be easily obtained by taking 
$\Omega{\cal R}$ invariant combinations. There are several 
sectors to be taken into account \cite{bgkl}:

\begin{itemize}
\item $D_a-D_a$: the $\Omega{\cal R}$ takes this sector 
to the $\Omega{\cal R}D_a-\Omega{\cal R}D_a$ sector. 
In general these sectors will be different and one should 
only take one of these into account. 
This sector contains, as in the toroidal case, d=4 
$\cn = 4$ super Yang-Mills. When one brane is its own orientifold
image $SO(N)$ and $USp(N)$ groups 
can appear (See, for instance, \cite{bgkl}.). 
As we are interested in unitary groups we will not consider these cases.  
  
\item $D_a-D_b$: the $\Omega{\cal R}$ takes this sector 
to the $\Omega{\cal R}D_b-\Omega{\cal R}D_a$ sector. 
One obtains chiral 
fermions in the bifundamental $(N_a,\bar{N}_b)$ of the group
with multiplicity given by the intersection number :
\beq
I_{ab}\ =\
(n_a^1m_b^1-m_a^1n_b^1)(n_a^2m_b^2-m_a^2n_b^2)(n_a^3m_b^3-m_a^3n_b^3)
\label{internumber}
\eeq
\item $D_a-\Omega{\cal R}D_b$: this sector is taken to the 
$D_b-\Omega{\cal R}D_a$ sector. There are  chiral fermions in the 
$(N_a,N_b)$ representation with multiplicity
\beq
I_{ab^*}\ =\
-(n_a^1m_b^1+m_a^1n_b^1)(n_a^2m_b^2+m_a^2n_b^2)(n_a^3m_b^3+m_a^3n_b^3)\ .
\label{internumber2}
\eeq
In eqs.(\ref{internumber}) and (\ref{internumber2}) a negative sign implies
opposite chirality.

\item $D_a-\Omega{\cal R}D_a$: this sector is taken to 
the $D_a-\Omega{\cal R}D_a$ sector. This is an invariant 
sector so the $\Omega{\cal R}$ projection should be imposed 
on it: some of the intersections will be invariant and the others will be in pairs.
 The invariant ones give $8 m_a^1 m_a^2 m_a^3$ fermions in
the antisymmetric 
representation and the others produce  $4 m_a^1 m_a^2 m_a^3 (n_a^1 n_a^2
n_a^3 -1)$ 
symmetric and antisymmetric representations \cite{bgkl}
\footnote{Notice that when $\prod_{i=1}^{3} n_a^i= 0$ we 
still have a $U(N_a)$ gauge group 
with chiral fermions living on it. 
In general there will be $4 \prod_{i=1}^{3} m_a^i$ fermions 
in the antisymmetric and the same number of 
fermions in the symmetric, but with opposite 
chirality. This will give us the same 
contribution to chiral  $SU(N_a)$ anomalies as the general formula.
This system is analogous to some constructions of non-BPS D-branes of Type I theory (see \cite{D7}).}.

\end{itemize}

RR tadpole conditions can be easily obtained from 
the toroidal case taking into account that some of 
the conditions are immediately satisfied because 
the pairs of branes cancel the contribution to some 
cycle conditions (the cycles with an odd number of $m_i$'s). 
The orientifold plane introduces a net RR charge 
in the $(1,0),(1,0),(1,0)$ cycle. So the tadpole conditions read 

\begin{eqnarray}
\sum_a N_a n_a^1 n_a^2 n_a^3 = 16 \nonumber\\
\sum_a N_a m_a^1 m_a^2 n_a^3 = 0 \nonumber\\
\sum_a N_a m_a^1 n_a^2 m_a^3 = 0 \nonumber\\
\sum_a N_a n_a^1 m_a^2 m_a^3 = 0
\end{eqnarray}
%
One  can also consider the possibility of 
adding a NS B-flux \cite{bfield}, $b^i$, to each two dimensional torus in
the  D9-brane picture \cite{bkl}. The total flux in the brane
becomes a  combination of the magnetic and B-field flux,
${\cal F} = b + F$.
In the T-dual picture the torus changes 
its complex structure in such a way that it 
takes into account the modified angle of the brane
${\cal F} = \cot{\vartheta} $.
Notice that the B-field is not invariant under $\Omega$. 
It is not a dynamical field but some discrete values are allowed: 
$b = 0,1/2$. In the T-dual picture the B-field is translated 
into a fixed complex structure \cite{bkl}. In an effective manner
the addition of this B-background allows for 
semi-integer  $m^i$ values. If originally the wrapping numbers 
on a torus are $(n,m')$, the effective wrapping numbers 
upon the addition of a $b=1/2$ background are $(n,m'+n/2)$
\cite{bkl}.
In what follows we will denote by $m=m'+n/2$ in those tori
with a B-background.

\subsection{$U(1)$ Anomaly cancellation}

Anomaly cancellation of $U(1)$'s for toroidal models were 
already considered in ref.\cite{afiru}. In the orientifold case
here considered there are some simplifications compared to the toroidal
case. Let us  consider the T-dual version consisting of Type I
string theory (D9-branes) with magnetic fluxes. We have in ten
dimensions RR fields $C_2$ and $C_6$ only with world-volume
couplings (wedge products are understood) :
\beq
\int_{D9_a} C_6 \, F_a^2  \ ;\  \int_{D9_a} C_2 \, F_a^4 \ . 
\label{rrcoup}
\eeq
Upon dimensional reduction we get one two form 
plus three other two-forms:
\beqa
B_2^0 & = & C_2 \nonumber \\ 
B_2^{I} &  =  & \int_{(T^2)_J\times (T^2)_K} C_6  \ ; I=1,2,3
\label{beillos}
\eeqa
with $I\not=J\not=K\not=I$ and their four-dimensional duals:
\beqa
C^0 & = & \int_{(T^2)_1\times (T^2)_2\times (T^2)_3} C_6 \nonumber\\
C^{I} & = & \int_{(T^2)_I} C_2  \
\label{ceillos}
\eeqa
with $dC^0  = -  * d B_2^0$ and $dC^I  =  - * d B_2^I$.  
These RR fields have four-dimensional couplings to the gauge fields
given by \cite{afiru}:
\beqa
\begin{array}{ccc}
N_a\, m^1_a\, m^2_a\, m^3_a \int_{M_4} B_2^0 \wedge F_a & \quad ; \quad
& n^1_b\, n^2_b\, n^3_b \int_{M_4} C^0 \wedge F_b \wedge F_b \nonumber \\
N_a\, n^J_a\, n^K_a\, m^I_a \int_{M_4} B_2^I \wedge F_a & \quad ; \quad
& n^I_b\, m^J_b\, m^K_b \int_{M_4} C^I \wedge F_b\wedge F_b  \ .  
\end{array}
\eeqa 
The Green-Schwarz amplitude where $U(1)_a$ couples to one of the
$B_2$ RR fields which propagates and couples to two $SU(N_b)$ 
gauge bosons will be proportional to:
\beq
-N_a\,\; m^1_a m^2_a m^3_a n^1_b n^2_b n^3_b \ -\  
N_a\, \sum_I n^I_a n^J_a m^K_a n^K_b m^I_b m^J_b  \ ,\ I\not= J,K 
\label{contgs}
\eeq
which precisely has the form to cancel the triangle anomalies.
Notice that due to the linear couplings between the $U(1)$'s and the 
RR two-forms some of the $U(1)$'s (including all those which are
anomalous) will become massive. Since there are only four 
two-forms available, in any model with any arbitrary number
of branes only a maximum of four $U(1)$'s may become massive.
Notice also that in any realistic model we will have to ensure that the
physical hypercharge generator is not one of them.

\section{Searching for the standard model}

Let us try to construct a specific model with low-energy spectrum given by
that in table 1, corresponding to the intersection numbers in
eq.(\ref{intersec2}). We find that getting the spectrum of the SM
is quite a strong constraint.  We find  families of models
corresponding to
choices of wrapping numbers $n_i^l$, $m_i^l$, $i=a,b,c,d$, $l=1,2,3$ as well
 as adding a NS $B$-background or not on  the three underlying tori. To
motivate the form
of these solutions let us enumerate some of the constraints
we have to impose:

{\bf 1)}
We will require that for any brane $i$ one has
$\Pi_{l=1}^3\ m_i^l\ =0$ because of two reasons. First, to
avoid the appearance of matter at the intersections of
one brane to its mirror. This matter (transforming like
symmetric or antisymmetric representations of the gauge group)
has exotic quantum numbers beyond the particle content of the
SM which we are trying to reproduce. In addition, there
are tachyonic scalars at those intersections which would destabilize
the brane configuration.

{\bf 2)}
If $\Pi_{l=1}^3\ m_i^l\ =0$ is verified, then in these toroidal models
there are only at most three RR fields $B^l$, $l=1,2,3$ with couplings
to the Abelian groups. Thus at most three $U(1)$'s may become
massive by the mechanism described in chapter 2. This implies
that we should consider only models with four sets of branes at
most, since otherwise there would be additional massless $U(1)$
gauge bosons beyond hypercharge.

{\bf 3)} We further impose that we should reproduce the spectrum in
table 1. This is the most constraining condition. It implies
that the branes $a$ should be parallel to the $d$ brane
in at least one  of the three complex planes and that the
$b$ branes are parallel to the $c$ brane. Getting $I_{ab}=1$ and
$I_{ab^*}=2$ requires that at least one of the three tori (e.g.,the third)
should be tilted (or should have a b-background, in the T-dual
language). Getting the other intersection numbers correct
gives us also further constraints.

Imposing these conditions we find the  general class of
solutions for the wrapping numbers shown in table (\ref{solution}).

\begin{table}[htb] \footnotesize
\renewcommand{\arraystretch}{2.5}
\begin{center}
\begin{tabular}{|c||c|c|c|}
\hline
 $N_i$    &  $(n_i^1,m_i^1)$  &  $(n_i^2,m_i^2)$   & $(n_i^3,m_i^3)$ \\
\hline\hline $N_a=3$ & $(1/\beta ^1,0)$  &  $(n_a^2,\epsilon \beta^2)$ &
 $(1/\rho ,  1/2)$  \\
\hline $N_b=2$ &   $(n_b^1,-\epsilon \beta^1)$    &  $ (1/\beta^2,0)$  &  
$(1,3\rho /2)$   \\
\hline $N_c=1$ & $(n_c^1,3\rho \epsilon \beta^1)$  & 
 $(1/\beta^2,0)$  & $(0,1)$  \\
\hline $N_d=1$ &   $(1/\beta^1,0)$    &  $(n_d^2,-\beta^2\epsilon/\rho )$  &  
$(1, 3\rho /2)$   \\
\hline \end{tabular}
\end{center} \caption{ D6-brane wrapping numbers giving rise to a SM spectrum.
The general solutions 
 are parametrized by a phase $\epsilon =\pm1$, the NS background
on the first two tori $\beta^i=1-b^i=1,1/2$, four integers
$n_a^2,n_b^1,n_c^1,n_d^2$ and a parameter $\rho=1,1/3$.
In order to obtain the correct hypercharge massless $U(1)$ 
those parameters have to verify the extra constraint
eq.(\ref{condhiper}).
\label{solution} }
\end{table}

In the table we have 
 $\beta^i =1-b^i$, with $b^i=0,1/2$  being the NS B-background
field discussed in  section 3. In the third torus one always has 
$b^3=1/2$.  Also $\epsilon=\pm 1$ and 
$\rho$ takes only the values $\rho=1,1/3$.
Notice that each of these families of D6-brane configurations depends on
four integers ($n_a^2,n_b^1,n_c^1$ and $n_d^2$)
\footnote{Care should be taken when choosing these integers
to have well-defined wrapping numbers in our tilted tori. If, for instance,
$\beta^1=1/2$, then $n_b^1,n_c^1$ should be odd integers, same with 
$\beta^2=1/2$ and $n_a^2,n_d^2$. By the same token, if we only want to 
consider this minimal gauge group we should consider coprime wrapping
numbers, so if $\rho=1$  then $n_c^1$ cannot be a multiple of 3, etc.}.
All of the choices lead exactly to the same massless
 fermion spectrum of table 1.

One has now to ensure that these choices are consistent with the
tadpole cancellation conditions described in the previous section.
It turns out that all but the first of those conditions are
automatically satisfied by the above families of configurations.
The first tadpole condition reads in the present case:
\beq
\frac{3n_a^2}{\rho \beta^1} \ +\ \frac{2n_b^1}{\beta^2} \ +\
\frac{n_d^2}{\beta^1} \ = \
16 \ .
\label{tadsm}
\eeq
Note  however that one can always relax this constraint by
  adding  extra D6-branes with no 
intersection with the SM ones and not contributing to the 
rest of the tadpole conditions. For example, a simple  possibility
would be the addition $N_h$ D6 branes with $m_h^l=0$, 
i.e., parallel to the orientifold plane. In this case the above
condition would be replaced by the more general one
\beq
\frac{3n_a^2}{\rho \beta^1} \ +\ \frac{2n_b^1}{\beta^2} \ +\
\frac{n_d^2}{\beta^1} \ +\ N_hn_h^1n_h^2n_h^3\  = \
16 \ .
\label{tadsm2}
\eeq
Thus the families of standard model configurations we have found 
are very weakly constrained by tadpole cancellation conditions.
This is not so surprising. Tadpole cancellation conditions 
are closely connected to cancellation of anomalies. Since
 the SM is anomaly-free, it is not surprising that the solutions we
find almost automatically are tadpole-free.

Let us now analyze the general structure of $U(1)$ anomaly
cancellation in this class of models.
As we remarked in section 2, there are two anomalous 
$U(1)$'s given by the generators $(3Q_a-Q_d)$ and $Q_b$
and two anomaly free ones which are $(Q_a+3Q_d)$ and $Q_c$.
Following the general discussion in previous section 
one can see that the three RR fields $B_2^I$, $l=1,2,3$ 
couple to the $U(1)$'s in the models as follows:
\beqa
B_2^1 &\wedge & \ {{-2\epsilon \beta^1}\over { \beta^2 } }F^{b} \nonumber \\
B_2^2 &\wedge  & \ \frac{(\epsilon  \beta^2 )}
{\rho \beta^1}(3F^a\ -\  F^{d})
\nonumber \\
 B_2^3 &\wedge  &  \ {1\over {2 \beta^2 } }
(\frac{3\beta^2 n_a^2}{\beta^1} F^a\ +\ 6\rho n_b^1F^{b} \ +\ 
2n_c^1F^c \ +\ \frac{3\rho  \beta^2 n_d^2}{\beta^1} F^d) \
\label{bfs}
\eeqa
whereas the $B_2^0$ RR field has no couplings to the $F_j$, because
$\Pi_lm_j^l=0$ for all the branes. The dual scalars $C^I$ and
$C^0$ have couplings:
\beqa
 C^1\ & (\frac{\epsilon \beta^2}{2\beta^1})& (F^a\wedge F^a\ -\ 3F^d\wedge
F^d)\nonumber \\
 C^2\ &  (\frac{3\rho\epsilon\beta^1}{2\beta^2}) & (-F^b\wedge
F^b\ +\
2F^c\wedge
F^c)\nonumber
\\
  &  C^0     &  
(\frac{n_a^2}{\rho \beta^1} F^a\wedge F^a
\ +\ \frac{n_b^1}{\beta^2} F^b\wedge F^b \ +\ \frac{n_d^2}{\beta^1}
F^d\wedge F^d) \
\label{cff}
\eeqa
and the RR scalar $C^3$ does not couple to any $F\wedge F$ term.
It is easy to check that these terms cancel all residual
$U(1)$ anomalies in the way described in section 2. Notice
in particular how only the exchange of the $B_2^1,B_2^2$ fields
(and their duals $C^1,C^2$) can contribute to the cancellation 
of anomalies since the $C^3$ field does not couple to $F\wedge F$
and $B_2^0$ does not couple to any $F^j$.
The exchange of those RR fields  proceeds in a universal manner
(i.e., independent of the particular choice of  $n$'s)
and hence the mechanism for the $U(1)$ anomalies to cancel 
is also universal. 
 On the other hand 
the $B_2^3$ field does couple to a linear combination 
of the four $U(1)$'s and hence will render that combination massive.
The $U(1)$ which remains light is given by the linear combination
\footnote{In the particular case with $n_c^1=0$ one can have 
both anomaly-free $U(1)$'s remaining in the massless spectrum
as long as one also has $n_a^2=n_d^2=0$.}
\beq
Q\ =\ n_c^1(Q_a+3Q_d) \ -\ {{3\beta ^2}\over {2\beta^1}}
(n_a^2+3\rho n_d^2)Q_c
\label{unolig}
\eeq
If we want to have just the standard hypercharge at low energies this
should be proportional to the hypercharge generator. This is the case
as long as:
\beq
n_c^1 \ =\ {{\beta^2}\over {2\beta^1} } (n_a^2+3\rho n_d^2)
\label{condhiper}
\eeq
which is an extra condition the four integers should fulfill 
in order to really obtain a SM at low energies.
Thus we have found families of toroidal models with 
D6-branes wrapping at angles in which the residual gauge group
is just the standard model $SU(3)\times SU(2)\times U(1)_Y$
and with three standard generations of quarks and leptons and
no extra chiral fermions (except for three right-handed 
neutrinos which are singlets under hypercharge).

These models are specific examples of the general approach in
section 2. Notice in particular that in these models 
Baryon number ($Q_a)$, and  lepton number ($Q_d$) are 
unbroken gauged  $U(1)$ symmetries. The same is the case of the
symmetry $Q_b$ which is a (generation dependent) Peccei-Quinn 
symmetry. Once the RR-fields give masses to three of the 
$U(1)$'s of the models, the corresponding $U(1)$'s 
remain as effective global symmetries in the theory.
This has the important physical consequences:

1) Baryon number is an exact perturbative symmetry of the
effective Lagrangian. Thus the nucleon should be stable.
This is a very interesting property which is quite a general
consequence of the structure of the theory in terms of 
D-branes intersecting at angles and which was already advanced in
ref.\cite{afiru2}.  Notice that this property is particularly welcome in 
brane scenarios with a low energy string scale \cite{aadd,otherbw} in which 
stability of the proton is an outstanding difficulty.
But it is also a problem in standard scenarios like the MSSM 
in which one has to impose by hand discrete symmetries like R-parity
or generalizations in order to have a sufficiently stable proton.

2) Lepton number is an exact symmetry in perturbation theory.
This has as a consequence that Majorana masses for the 
neutrinos should be absent. Any neutrino mass should be
of standard Dirac type. They can however be naturally small 
as we discuss below.

3) There is a gauged $U(1)$ symmetry of the Peccei-Quinn type
($Q_b$) which is exact at this level. Thus, at this level
 the $\theta_{QCD}$ parameter can be  rotated to zero.

These properties seem to be quite model independent, and also seem to
be a generic property of any D-brane model which gives rise
to {\it just} a SM spectrum at the intersections.

As a final comment note that the pseudoscalar $C^0$ remains
massless at this level 
and has axionic couplings (eq.(\ref{cff})) to the gauge fields of the 
SM (and also to the fields coming from the extra branes added
to cancel tadpoles, if present). It would be interesting to study the
possible relevance of this axion-like field concerning
the strong CP problem.

\section{Absence of tachyons and stability of the configurations }

\begin{figure}
\centering   
\epsfxsize=3in
\hspace*{0in}\vspace*{.2in}
\epsffile{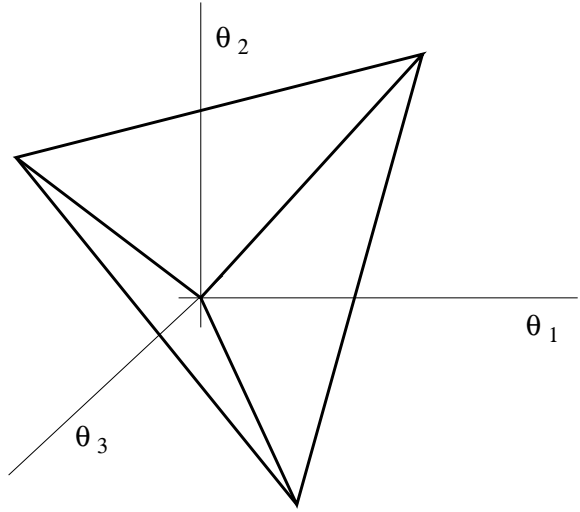}
\caption{\small 
The region inside the tetrahedron has no tachyons.
Faces, edges and vertices represent
respectively, $\cn = 1$, $\cn = 2$ and  $\cn = 4$ systems
at the given intersection.}
\label{tetra}
\end{figure}

We have been concerned up to now with the massless 
chiral fermions at the D-brane intersections. 
In addition to those
there are scalar states at each intersection
which in some sense may be considered 
(in a sense specified below)  "SUSY-partners", squarks and sleptons,
of the massless chiral fermions, since they have the same 
multiplicity $|I_{ij}|$  and carry the same gauge quantum numbers
\footnote{Notice that these masses are the same for all intersections corresponding
to the same pair of branes. This flavour independence is interesting
from the point of view of suppression of flavour-changing
neutral currents.}.
The lightest of those states have  masses
\cite{afiru}
{\small \beqa
\begin{array}{cc}
{\rm \bf State} \quad & \quad {\bf Mass^2} \\
t_1=(-1+\vartheta_1,\vartheta_2,\vartheta_3,0) & \alpha' {\rm (Mass)}^2 =
\frac 12(-\vartheta_1+\vartheta_2+\vartheta_3) \\
t_2=(\vartheta_1,-1+\vartheta_2,\vartheta_3,0) & \alpha' {\rm (Mass)}^2 =
\frac 12(\vartheta_1-\vartheta_2+\vartheta_3) \\
t_3=(\vartheta_1,\vartheta_2,-1+\vartheta_3,0) & \alpha' {\rm (Mass)}^2 =
\frac 12(\vartheta_1+\vartheta_2-\vartheta_3) \\
t_4=(-1+\vartheta_1,-1+\vartheta_2,-1+\vartheta_3,0) & \alpha' {\rm (Mass)}^2
= 1-\frac 12(\vartheta_1+\vartheta_2+\vartheta_3)
\label{tachdsix}
\end{array}
\eeqa}
in the notation of ref.\cite{afiru}.
Here $\vartheta_i $ are the intersection angles (in units of $\pi $)
at each of the three subtori. As is obvious
from these formulae  the masses depend on the angles at each intersection
and hence on the relative size of the radii.
Thus in  principle some of the scalars may be tachyonic 
\footnote{One can check that for models
with positive $n^i$ the scalar $t_4$ can never become tachyonic.}.
In fig.\ref{tetra} we show  the range of  $\vartheta_i $ for which
there are no tachyons at a given intersection. 
 There is a region (inside the tetrahedron) where all the scalars have 
positive $(mass)^2$. Supersymmetry is not preserved
but the absence of tachyons indicates that
the system cannot decay into another one that lowers the energy.
 Outside this region some scalars 
become tachyonic.  
 The boundary between the two
regions represents a supersymmetric
configuration at that intersection.
This boundary has a tetrahedral shape.
The faces represent configurations that
preserve $\cn = 1$, the edges correspond to $\cn = 2$ configurations
and the vertices to $\cn = 4$ configurations at that 
particular intersection. At each of the faces a different 
scalar becomes massless
and hence becomes degenerate with the chiral fermion in
the intersection. One can check that if none of
the other scalars is tachyonic there is a fermion-boson
degeneration that indicates that one supersymmetry is preserved
locally. It is in this sense that these scalars are 
SUSY-partners of the massless chiral fermion. At the edges 
it is two scalars (and one fermion) which become massless 
and one has (locally) $\cn = 2$ supersymmetry.

\begin{figure}
\centering
\epsfxsize=6in
\hspace*{0in}\vspace*{.2in}
\epsffile{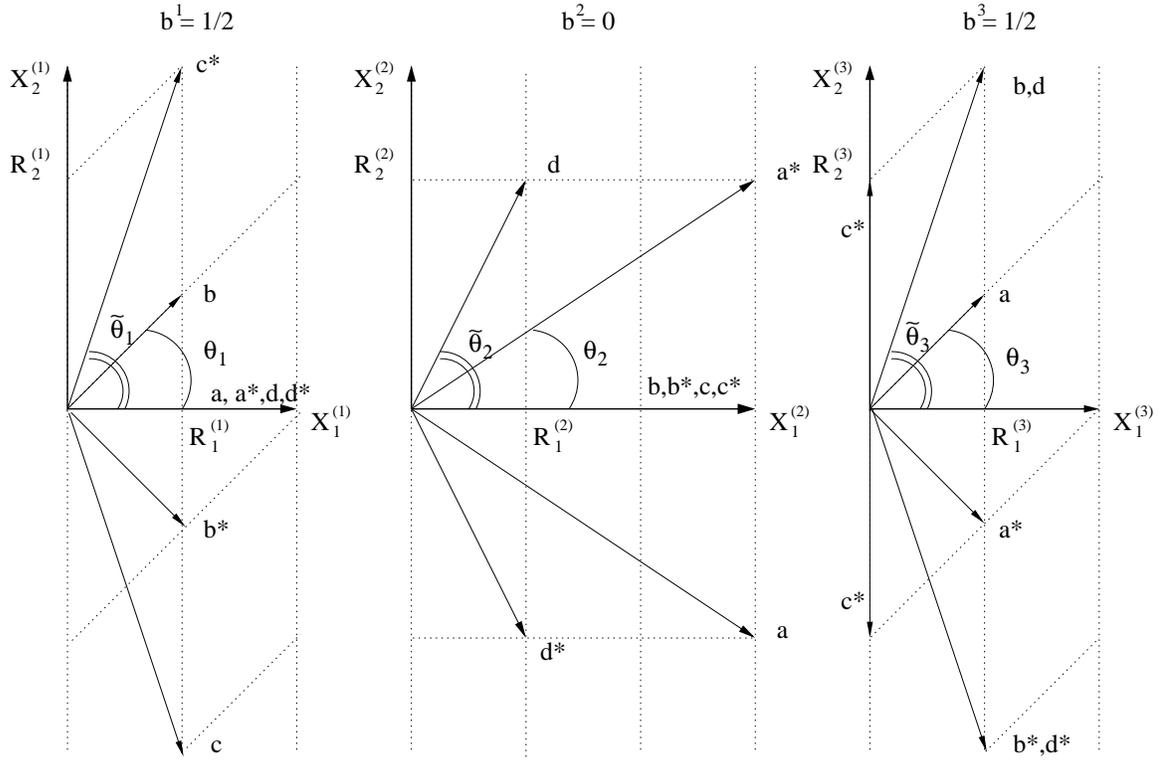}
\caption{\small  Definition of the angles between the different branes 
on the three tori. We have selected a particular setting where $n_a^2,n_b^1,n_c^1,n_d^2 > 0$, 
$\epsilon = -1$ and $\beta^1 = 1/2$, $\beta^2= 1$.}
\label{angles}
\end{figure}

For a D-brane configuration to be stable there should be no tachyons 
at {\it none } of the intersections. As 
already noted in ref.\cite{afiru}, in general it is possible 
to vary the compact radii in order  to get rid of all tachyons
of a given model. One can do a general analysis of sufficient 
conditions for absence of tachyons in the standard model
examples of previous sections which are parametrized in terms 
of $\beta^{1,2}$ and the integers $n_a^2,n_b^1,n_c^1$ and $n_d^2$.
Let us define the angles 
\beqa
\theta_1 \   = \ \frac{1}{\pi} cot^{-1}\frac{n_b^1R_1^{(1)}}{\beta^1R_2^{(1)}} \ ;\
\theta_2 \  =   \  \frac{1}{\pi} cot^{-1}\frac{n_a^2R_1^{(2)}}{\beta^2R_2^{(2)}} \ ;\
\theta_3 \  = \  \frac{1}{\pi} cot^{-1}\frac{2R_1^{(3)}}{\rho R_2^{(3)}} \nonumber \\
{\tilde {\theta_1}} \   = 
\ \frac{1}{\pi} cot^{-1}\frac{|n_c^1| R_1^{(1)}}{3\rho \beta^1 R_2^{(1)}}\ ;\
{\tilde { \theta_2}} \ 
 = \ \frac{1}{\pi} cot^{-1}\frac{\rho n_d^2R_1^{(2)}}{\beta^2R_2^{(2)}}   \ ;\
{\tilde {\theta_3 }} \  = \ \frac{1}{\pi} cot^{-1}\frac{2R_1^{(3)}}{3\rho R_2^{(3)}}
\label{angulos}
\eeqa
where $R^{(i)}_{1,2}$ are the compactification radii
for the three $i=1,2,3$ tori
\footnote{As can be seen in
 fig.(\ref{angles}), $R^{(i)}_{1}$ are not compactification 
radii in a strict sense 
if $b^i \neq 0$ but their projection onto the $X^{(i)}_{1}$ direction.}.
The geometrical meaning of the angles is depicted in fig.(\ref{angles}).

 Angles at all the intersections may be written in terms of 
those six angles which depend on the parameters of the
particular model and the relative radii. 
We have four (possibly light) scalars $t_i, i=1,2,3,4$ at each of the 
7 independent types of intersections, thus altogether 28 
different scalar masses. Since all these 28 masses 
can be written in terms of the above 6 angles, it is obvious that
the masses are not all independent.
Thus for example one finds:
\beqa
m^2_{ab}(t_2)+m^2_{ac}(t_3) \ =\ m^2_{ab*}(t_2)+m^2_{ac*}(t_3)
\ =\ m^2_{bd*}(t_2)+m^2_{cd*}(t_3) \nonumber \\
m^2_{ab}(t_1)+m^2_{ac}(t_4) \ =\ m^2_{ab*}(t_1)+m^2_{ac*}(t_4)
\ =\ m^2_{bd*}(t_1)+m^2_{cd*}(t_4) \nonumber \\
m^2_{ab}(t_1)+m^2_{ac*}(t_2) \ =\ m^2_{ab*}(t_1)+m^2_{ac}(t_2)
\ =\ m^2_{bd*}(t_2)+m^2_{cd}(t_1) \nonumber \\
m^2_{ab}(t_2)+m^2_{ac*}(t_1) \ =\ m^2_{ab*}(t_2)+m^2_{ac}(t_1)
\ =\ m^2_{bd*}(t_1)+m^2_{cd}(t_2)  \
\label{sumrules}
\eeqa
These give interesting relationships among the squark and 
slepton partners of usual fermions.
Due to these kind of constraints 
 the  28 conditions
for absence of tachyons may be reduced to only
14 general conditions (see Appendix II).

 In order to get an idea of how easy is to get
a tachyon-free configuration in one of the standard model
examples of the previous section let us consider a particular case.
Consider a model  with $\rho =\beta^1=\beta^2=1$,
$\epsilon=-1$ and with $n_a^2=2$,
$n_b^1=n_d^2=0$ and 
$n_c^1=1$. The wrapping numbers of the four stacks of branes 
are thus:
\beqa
N_a=3 \ &\ (1,0)  (2,-1) (1,1/2)  \nonumber \\
N_b=2 \ &\ (0,1)  (1,0) (1,3/2)  \nonumber \\
N_c=1 \ &\ (1,-3)  (1,0) (0,1)  \nonumber \\
N_d=1 \ &\ (1,0)  (0,1) (1,3/2)    
\label{solu2} 
\eeqa
This verifies all the conditions to get just the
SM gauge group with three quark/lepton generations. 
The first tadpole condition may be fulfilled by
adding e.g. 5  parallel branes with $n^1=n^2=1$, $n^3=2$ and
$m^i=0$. Now, in this case one has $\theta_1 = 1/2 >
{\tilde {\theta_1}}$, $\theta_2 = 1/2$ and many of the 
equations shown in the Appendix II are trivially satisfied.
 Then one can check that 
there are no tachyons at {\it  any } of the intersections
as long as:
\beqa
\theta _2 \ +\ {\tilde {\theta_3}} \ -\ \theta_3 \ & \geq & \frac{1}{2}
\nonumber \\
\tilde{\theta_1} \ \geq \ {\tilde {\theta_3}}
\label{condi}
\eeqa
which may be easily satisfied for wide ranges of the radii.
Similar simple expressions are obtained in other examples.
For instance, a model within the first family in Table 4
with $n_a^2=0,n_b^1=-1,n_c^1=1,n_d^2=1$ and 
$\rho=1/3$, $\beta^1=1/2,\beta^2=1$ has no tachyons as long as
the two conditions 
${\tilde \theta}_1+{\tilde \theta}_3-\theta_3\geq  1/2$  and
${\tilde \theta}_1+{\tilde \theta}_2-{\tilde \theta_3}\geq 1/2$
are verified. Again, this happens for wide ranges of the
radii.

\section{Spectrum of massive particles beyond the SM}

The open string spectrum consists of open strings stretched between
the different sets of D-branes (see fig.(\ref{sectors})).
The spectrum can be split into two sets: 

\begin{figure}
\centering
\epsfxsize=5in
\hspace*{0in}\vspace*{.2in}
\epsffile{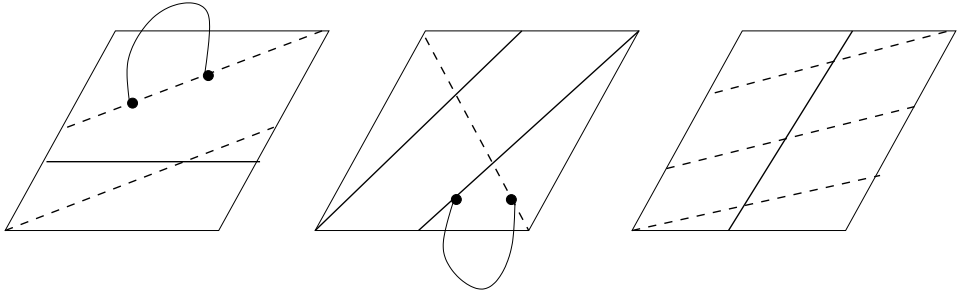}
\caption{\small Each rectangle represents a two torus.
There are two branes: one is represented by a straight
black line and the other by a dashed line. 
The curved lines represent strings ending on the D-branes.}
\label{sectors}
\end{figure}   

\begin{itemize}

\item  $D6_a-D6_b$ sector: strings ending on different
sets of $D6$-branes. The massless spectrum
 consists of $|I_{ab}|$ chiral fermions where
the chirality is determined by the sign of the
intersection number. In our case these are the standard
quarks and leptons which we have analyzed above. At those intersections
live also  the massive scalars we have described in the previous section
which in some sense will be SUSY-partners, squarks and sleptons,
of the ordinary particles.

There are also additional string excitations \cite{bdl}
which may be
relatively light depending on the angles
(these are the gonions of \cite{afiru2}). The mass
gap will be proportional to the angles between the
branes, $\vartheta_{ab}$, on each torus. These gonion
states include vector  boson and fermion massive replicas.
Here we just describe the lightest ones.
 In particular there are fermionic
states of the form:
{\small \beqa
\begin{array}{cc}   
{\rm \bf State} \quad & \quad {\bf Mass^2} \\
(1/2+\vartheta_1,-1/2+\vartheta_2,-1/2+ \vartheta_3,-1/2)
 & \alpha' {\rm (Mass)}^2 = \ \vartheta_1  \\
(-1/2+\vartheta_1,1/2+\vartheta_2,-1/2+ \vartheta_3,-1/2)
 & \alpha' {\rm (Mass)}^2 = \ \vartheta_2  \\
(-1/2+\vartheta_1,-1/2+\vartheta_2,1/2+ \vartheta_3,-1/2)
 & \alpha' {\rm (Mass)}^2 = \ \vartheta_3 \\
(-3/2+\vartheta_1,-1/2+\vartheta_2,-1/2+ \vartheta_3,-1/2)
 & \alpha' {\rm (Mass)}^2 = \ 1\ -\ \vartheta_1  \\
(-1/2+\vartheta_1,-3/2+\vartheta_2,-1/2+ \vartheta_3,-1/2) 
 & \alpha' {\rm (Mass)}^2 = \ 1\ -\ \vartheta_2  \\
(-1/2+\vartheta_1,-1/2+\vartheta_2,-3/2+ \vartheta_3,-1/2)
 & \alpha' {\rm (Mass)}^2 = \ 1\ -\ \vartheta_3
\label{gonifermions}
\end{array}
\eeqa}
and their chiral partners. They would be sort of massive
fermionic partners of quarks and leptons but may be
relatively light if some of the angles is sufficiently small.
In addition there are vector fields
{\small \beqa
\begin{array}{cc}
{\rm \bf State} \quad & \quad {\bf Mass^2} \\
(\vartheta_1,-1+\vartheta_2,-1+\vartheta_3,\pm 1) & \alpha' {\rm (Mass)}^2 =
 \vartheta_1+(1-r) \\
(-1+\vartheta_1,\vartheta_2,-1+\vartheta_3,\pm 1) & \alpha' {\rm (Mass)}^2 =
 \vartheta_2+(1-r) \\
(-1+\vartheta_1,-1+\vartheta_2,\vartheta_3,\pm 1) & \alpha' {\rm (Mass)}^2 =
\vartheta_3+(1-r)
\label{gonivectors}
\end{array}
\eeqa}
where $r=1/2(\vartheta_1 +\vartheta_2 +\vartheta_3)$.   
Finally there are extra scalars  beyond those described in the 
previous section: 
{\small \beqa
\begin{array}{cc}
{\rm \bf State} \quad & \quad {\bf Mass^2} \\
\alpha_{-\vartheta_i}
(-1+\vartheta_1,-1+\vartheta_2,-1+ \vartheta_3,0)
 & \alpha' {\rm (Mass)}^2 = \ \vartheta_i+(1-r)  \ ;\ i=1,2,3 \\
\alpha_{\vartheta_i -1}
(-1+\vartheta_1,-1+\vartheta_2,-1+ \vartheta_3,0)   
 & \alpha' {\rm (Mass)}^2 = \ (1-\vartheta_i)+(1-r)  \ ;\ i=1,2,3 \\
\label{goniscalars}
\end{array}
\eeqa}
These states are in general heavier than the scalars considered
in the previous section.
If some of the angles are small, further excitations appear from
acting with twisted oscillator operators ${\tilde {\alpha}}_{-\vartheta_i}$
and/or ${\tilde {\alpha}}_{\vartheta_i-1}$ on the above states.

Notice that, unlike the case of D4-branes discussed in ref.\cite{afiru2},
in the present case there is a priori 
no reason for any of the intersection angles of the 
configurations to be small and hence all the states considered in this
subsection may have masses of order the string scale.

\item $D6_a-D6_a$ sector: strings ending on the same set of
$D6$-branes.

In principle the 
 massless spectrum in this sector is just SYM $\cn = 4$
in four dimensions. However as explained in Appendix I, quantum effects
like those shown in fig.(\ref{oneloopmass}) will give masses to all particles in the
$\cn = 4$ multiplets except for the gauge bosons. Thus only the 
chiral SM fermions (and the SM gauge bosons) will remain at the 
massless level.

\begin{figure}
\centering
\epsfxsize=2in
\hspace*{0in}\vspace*{.2in}
\epsffile{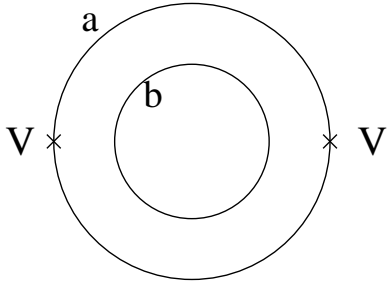}
\caption{\small One loop contribution to the masses of  the $\cn=4$ multiplet
states in the bulk of the branes}
\label{oneloopmass}
\end{figure}

In addition
 there are three types of massive
particles in this sector  \cite{afiru2}:

\begin{itemize}
\item 

For each stack of branes there will be 
KK excitations along the direction
where the D-brane is living. Their masses  are:

\beq
 m = \frac{k_1}{l_1} + \frac{k_2}{l_2}  + \frac{k_3}{l_3}
\eeq

where $k_i$ are integer numbers reflecting
the KK mode on the $i$th torus and $l_i$ is the length of the brane on this
torus.

\item String winding states along the transverse
directions to the brane. Their masses  are  of the form:

\beq
 \alpha' m = \frac{A_1}{l_1} + \frac{A_2}{l_2}  + \frac{A_3}{l_3}
\eeq

where $A_i$ is the area of the  $i$th  two dimensional torus.

\item String excitations with a mass gap of  $(\alpha')^{-\frac{1}{2}}$.

\end{itemize}

\end{itemize}

The closed string spectrum is just the Kaluza Klein
reduction of the ten dimensional Type IIA spectrum. None of the
supersymmetries is broken in the toroidal compactification.
So we expect a $\cn = 8$ $d = 4$ supergravity multiplet living in the bulk.
The fact that on the D-brane network supersymmetry is broken 
will however transmit supersymmetry breaking to the bulk
closed string sector at some level.

\section{The Higgs sector and electroweak symmetry breaking}

Up to now we have ignored the existence or not of the Higgs system
required for the breaking of the electroweak symmetry as well as
for giving masses to quarks and leptons. Looking at the $U(1)$ charges of
quarks and leptons in Table 1, we see that possible Higgs fields 
coupling to quarks come in four varieties 
with charges under $Q_b,Q_c$ and hypercharge given in  Table 3.
\begin{table}[htb] \footnotesize
\renewcommand{\arraystretch}{1.25}
\begin{center}
\begin{tabular}{|c|c|c|c|}
\hline  
 Higgs   &  $Q_b$  &  $Q_c$   & Y \\
\hline\hline $h_1$ & 1  &  -1 & 1/2  \\
\hline $h_2$ &   -1    &  1  &  -1/2   \\
\hline\hline $H_1$ & -1  &  -1 & 1/2  \\        
\hline $H_2$ &   1    &  1  &  -1/2   \\    
\hline \end{tabular}
\end{center} \caption{ Electroweak Higgs fields
\label{higgsses} }
\end{table}
Now, the question is whether for some configuration of the branes 
such  Higgs fields  appear in the light spectrum.
Indeed that is the case. The $U(2)$ branes ($b,b^*$) are parallel 
to the ($c,c^*$) branes along  the second torus and hence they do not intersect.
However there are open strings which stretch in between  both sets of branes
and which lead to  light scalars when the distance $Z_2$ in the second torus
is small. In particular there are the scalar states 
{\small \beqa
\begin{array}{cc}
{\rm \bf State} \quad & \quad {\bf Mass^2} \\
(-1+\vartheta_1, 0, \vartheta_3, 0) & \alpha' {\rm (Mass)}^2 =
  { {Z_2}\over {4\pi ^2}}\ +\ \frac{1}{2}(\vartheta_3 - \vartheta_1) \\
(\vartheta_1, 0, -1+ \vartheta_3, 0) & \alpha' {\rm (Mass)}^2 =
  { {Z_2}\over {4\pi ^2 }}\ +\ \frac{1}{2}(\vartheta_1 - \vartheta_3) \\
\label{Higgsmasses}
\end{array}
\eeqa}
where $Z_2$ is the distance$^2$ (in $\alpha '$ units) 
 in transverse space along the second torus.
$\vartheta_1 $ and $\vartheta_3$ are the relative angles between the
$b$ and $c$ (or $b$ and $c^*$) in the first and third complex planes.
These four scalars have precisely the quantum numbers of the 
Higgs fields $H_i$ and $h_i$ in the table. The $H_i$'s come from 
the $b-c^*$ intersections whereas the $h_i$ come from the 
$b-c$ intersections. In addition to these scalars there are
two fermionic partners at each of $bc$ and $bc^*$ intersections
{\small \beqa
\begin{array}{cc}
{\rm \bf State} \quad & \quad {\bf Mass^2} \\
(-1/2+\vartheta_1, \mp 1/2 , -1/2+\vartheta_3, \pm 1/2 ) &  {\rm (Mass)}^2 =
  { {Z_2}\over {4\pi ^2\alpha '}}\  \\
\label{Higgsinomasses} 
\end{array}
\eeqa}
This Higgs system may be understood as  
massive $\cn=2$ Hypermultiplets containing respectively the $h_i$ and $H_i$ 
scalars along with the above fermions. The above scalar spectrum corresponds
to the following mass terms in the effective potential:
\beqa
V_2\ =\ m_H^2 (|H_1|^2+|H_2|^2)\ +\ m_h^2 (|h_1|^2+|h_2|^2)\ + \nonumber \\
+\ m_B^2 H_1H_2+h.c. \ +\ m_b^2h_1h_2+h.c.
\label{Higgspot}
\eeqa
where:
\beqa
 {m_h}^2 \ =\ { {Z_2^{(bc)}}\over {4\pi ^2\alpha '}}\ & ; & \ 
{m_H}^2 \ =\ {{Z_2^{(bc^*)}}\over {4\pi ^2\alpha '}}\nonumber \\
m_b^2\ =\ \frac{1}{2\alpha '}|\vartheta_1^{(bc)}-\vartheta_3^{(bc)}| \ & ;&
m_B^2\ =\ \frac{1}{2\alpha '}|\vartheta_1^{(bc^*)}-\vartheta_3^{(bc^*)}|
\label{masillas}
\eeqa
Notice that each of the Higgs systems 
have a quadratic potential similar to that of the  
MSSM.
In fact one also expects the quartic potential to be identical to that
of the MSSM.
In our case the mass parameters of the potential have an
interesting  geometrical  interpretation in terms of the brane distances 
and intersection angles.

What are the sizes of the Higgs mass terms? The values of $m_H$ and $m_h$
are controlled by the distance between the  $b,c,c^*$ branes 
in the second torus. These values are in principle free parameters and hence one
can make these parameters arbitrarily small compared to the string scale
$M_s$. That is not the case 
of the $m_B^2,m_b^2$ parameters. We already mentioned that all 
scalar mass terms  depend on only 6 angles in this class of models.
This is also the case  here, one finds (using also eq.(\ref{sumrules})):
\beqa
m^2_{B1}\ =&m^2_{Q_L}(t_2)+m^2_{U_R}(t_3) \ =\ m^2_{q_L}(t_2)+m^2_{D_R}(t_3)
\ =\ m^2_{L}(t_2)+m^2_{N_R}(t_3) \nonumber \\
m^2_{B2}\ =&m^2_{Q_L}(t_1)+m^2_{U_R}(t_4) \ =\ m^2_{q_L}(t_1)+m^2_{D_R}(t_4)
\ =\ m^2_{L}(t_1)+m^2_{N_R}(t_4) \nonumber \\
m^2_{b1}\ =&m^2_{Q_L}(t_1)+m^2_{D_R}(t_2) \ =\ m^2_{q_L}(t_1)+m^2_{U_R}(t_2)
\ =\ m^2_{L}(t_1)+m^2_{E_R}(t_2) \nonumber \\
m^2_{b2}\ =&m^2_{Q_L}(t_2)+m^2_{D_R}(t_1) \ =\ m^2_{q_L}(t_2)+m^2_{U_R}(t_1)
\ =\ m^2_{L}(t_2)+m^2_{E_R}(t_1) \nonumber \\
&m^2_B\ = min\{m^2_{B1},m^2_{B2}\}\ ;\ \ m^2_b = min\{m^2_{b1},m^2_{b2}\}
\label{sumrules2}
\eeqa
Thus if one lowers the $m_{B,b}^2$ parameters, some other scalar partners
of quarks and leptons have also to be relatively light, and one cannot
lower $m^2_{B,b}$ below present limits of these kind of scalars at 
accelerators.

Notice however that if the geometry is such that one approximately has
$m^2_H\ =m^2_B$ (and/or $m^2_h\ =m^2_b$ ) there appear scalar
flat directions along $<H_1>=<H_2>$ ($<h_1>=<h_2>$) which may
give rise to electroweak symmetry breaking at a scale well below
the string scale. Obviously this requires the string scale to be 
not far above the weak scale, i.e., $M_s=1-few$ TeV since otherwise
substantial fine-tuning would be needed.  
 Let us also point out that the particular Higgs
coupling to the top-quark (either $h_1$ or $H_1$) will in general get
an additional one-loop  negative contribution to its mass$^2$ in the
usual way \cite{ir}.

Let us have a look now at the number of Higgs multiplets which may
appear in the class of toroidal models discussed in previous sections.
Notice first of all that the number $n_H$($n_h$) of Higgs sets
of type $H_i$($h_i$) are given by the number of times 
the branes $b$ intersect with the branes $c$($c^*$) in the first and
third tori:
\beq
n_h\ =\  {\beta ^1} |n_c^1+3\rho n_b^1| \ ;\
n_H\ =\ {\beta ^1} |n_c^1-3\rho n_b^1|
\label{numhiggs}
\eeq
The simplest Higgs structure is obtained in the following
cases:

\begin{itemize}

\item {\it  Higgs system of the MSSM}

From eq.(\ref{numhiggs}) one sees that the 
 minimal set of Higgs fields is obtained when 
either $n_H=1,n_h=0$  or $n_H=0, n_H=1$.
For both of those cases it is easy to check that, after imposing
the condition eq.(\ref{condhiper} ), one is left   
with  two families of models
with $\rho=1/3,\beta^1=1/2$ depending on a single integer $n_a^2$
and on $\beta^2$.
These solutions are shown in the first four rows of table 3. 
\begin{table}[htb] \footnotesize
\renewcommand{\arraystretch}{1.25}
\begin{center}
\begin{tabular}{|c|c|c|c|c|c|c|c|c|}
\hline
 Higgs    &  $\rho $  &  $\beta^1$   & $\beta^2$ & $n_a^2$ & $n_b^1$ & $n_c^1 $
& $n_d^2$  & $N_h$ \\
\hline\hline  $n_H=1,n_h=0$  & 1/3  &  1/2  & $\beta^2$ & $n_a^2$ & -1 & 1& 
$\frac{1}{\beta^2}-n_a^2$   &  $4\beta^2(1-n_a^2)$  \\
\hline  $n_H=1,n_h=0$  & 1/3  &  1/2  & $\beta^2$ & $n_a^2$ & 1 & -1&        
$-\frac{1}{\beta^2}-n_a^2$   &  $4\beta^2(1-n_a^2)$  \\
\hline\hline  $n_H=0,n_h=1$  & 1/3  &  1/2  & $\beta^2$ & $n_a^2$ & 1 & 1&        
$\frac{1}{\beta^2}-n_a^2$   &  $4\beta^2(1-n_a^2)-1$  \\
\hline  $n_H=0,n_h=1$  & 1/3  &  1/2  & $\beta^2$ & $n_a^2$ & -1 & -1&        
$-\frac{1}{\beta^2}-n_a^2$   &  $4\beta^2(1-n_a^2)+1$  \\
\hline\hline  $n_H=1,n_h=1$  & 1   &  1   & $\beta^2$ & $n_a^2$ & 0  & 1&        
$\frac{1}{3}(\frac{2}{\beta^2}-n_a^2)$   &
$\beta^2(8-\frac{4n_a^2}{3})-\frac{1}{3}$ \\
\hline  $n_H=1,n_h=1$  & 1   &  1   & $\beta^2$ & $n_a^2$ & 0  & -1&
$\frac{1}{3}(-\frac{2}{\beta^2}-n_a^2)$   &
$\beta^2(8-\frac{4n_a^2}{3})+\frac{1}{3}$ \\
\hline  $n_H=1,n_h=1$  & 1/3   &  1   & $\beta^2$ & $n_a^2$ & 0  & 1&
$ \frac{2}{\beta^2}-n_a^2$   &
$\beta^2(8-{4n_a^2})-1$ \\
\hline  $n_H=1,n_h=1$  & 1/3   &  1   & $\beta^2$ & $n_a^2$ & 0  & -1&
$ -\frac{2}{\beta^2}-n_a^2$   &
$\beta^2(8-{4n_a^2})+1$ \\
\hline \end{tabular}
\end{center} \caption{ Families of models with the minimal Higgs content.
\label{minimal} }
\end{table}
The last column in the table shows the number $N_h$ of branes 
parallel to the orientifold plane one has to add in order to
cancel global RR tadpoles (a negative sign means  
antibranes).

As we will discuss in the following section, the minimal
choice with $n_H=1,n_h=0$ is particularly interesting 
\footnote{
It is amusing  that in this class of solutions with $n_a^2=1$ the SM sector
is already tadpole free and one does not need to add extra non-intersecting
 branes, i.e., $N_h=0$. Thus the SM is the only gauge group of the
whole model.}
from the
point of view of Yukawa couplings since the absence of the 
Higgs $h_i$ could be at the root of the smallness of 
neutrino masses. The opposite situation with $n_H=0$ and
$n_h=1$ is less interesting since charged leptons would
not get sufficiently large masses. For all the models of
the first family  with $n_H=1,n_h=0$   
the structure of the Higgs system of the three  models is analogous
and one gets:
\beqa 
 {m_H}^2 \ =\ { {(\xi_b+\xi_c)^2}\over {\alpha '}}\ ;\
m_B^2\ =\ \frac{1}{2\alpha '}|2\tilde\theta_1+ {\tilde {\theta}}_3-\frac{1}{2}|
\ ;
\label{masillas1}
\eeqa
where $\xi_b$($\xi_c$) is the distance between the orientifold plane
and the $b$($c$) branes and ${\tilde \theta}_1$, 
$ {\tilde {\theta}}_3$ were defined in
eq.(\ref{angulos}).

\item
{\it Double MSSM Higgs system}

The next to minimal set is having $n_H=n_h=1$.
After imposing the  condition eq.(\ref{condhiper} ) one finds 
 four families of such models depending on the integer $n_a^2$
and on $\beta^2$. They are shown in the last four rows of
table 3.  
The structure of the Higgs system in all these 4 families of
  models is
analogous and one gets:
\beqa
 {m_H}^2 \ =\ { {(\xi_b+\xi_c)^2}\over {\alpha '}}\ ;\
{m_h}^2 \ =\ {{(\xi_b-\xi_c)^2}\over {\alpha '}}\nonumber \\
m_B^2\ =\ \frac{1}{2\alpha '}|\tilde\theta_1+ {\tilde {\theta}}_3| \ ;
m_b^2\ =\ \frac{1}{2\alpha '}|\tilde\theta_1+ {\tilde {\theta}}_3|
\label{masillas2}
\eeqa

\end{itemize}

Let us finally comment that having a minimal set of Higgs fields
would automatically lead to absence of flavour-changing
neutral currents (FCNC) from higgs exchange. In the case
of a double Higgs system one would have to study in detail the structure
of Yukawa couplings in order to check whether FCNC are
sufficiently suppressed.

\section{ The  Yukawa and gauge coupling constants}

As we discussed in the previous section, 
there are four possible varieties of Higgs fields $h_i,H_i$ in this class
of models.
The Yukawa couplings among the SM fields in table 1 and the different
Higgs fields which are allowed  
 by the symmetries have the general form:
\beqa
y^U_jQ_LU_R^j h_1 \ +\ y^D_jQ_LD_R^jH_2 \ +  \nonumber \\
y^u_{ij}q_L^iU_R^j H_1 \ +\ y^d_{ij}q_L^iD_R^jh_2 \ + \nonumber \\ 
y^L_{ij}L^iE_R^jH_2  \ +\  y^N_{ij}L^iN_R^jh_1 \ +\  h.c. 
\label{yuki}
\eeqa
where $i=1,2$ and $j=1,2,3$. Which of the observed quarks 
(i.e. whether a given left-handed quark is inside 
$Q_L$ or $q_L^i$) fit into the multiplets will depend on 
which are the mass eigenstates of the quark and lepton mass
matrices after diagonalization. These matrices depend on the
Yukawa couplings in the above expression.   

The pattern of quark and lepton masses thus depends both
on the vevs of the Higgs fields $h_i,H_i$ and on the Yukawa
coupling constants and both dependences could be important
in order to understand the observed hierarchical structure. 
In particular it could be that e.g., only
one subset of the Higgs fields could get vevs. 
So let us consider two possibilities in turn.

\begin{itemize}

\item {\it Minimal set of Higgs fields}
This is for example
the case in the  situation with $n_H=1$, $n_h=0$ described
in the previous section in which only the $H_1,H_2$ fields 
appear. Looking at eq.(\ref{yuki}) we see that only
two $U$-quarks and  one $D$ quark  would get masses in this way. 
Thus one would identify them with the top, charm and b-quarks.
In addition there are also masses for charged leptons.
Thus, at this level, the  $s,d,u$-quarks would remain massless,
as well as the neutrinos. 

In fact this is not a bad starting point.
The reason why the $H_1,H_2$ fields do not couple to these other fermions
is because such couplings would violate the $U(1)_b$ symmetry
(see the table). On the other hand 
strong interaction effects will break such a symmetry 
and one expects that  they could  allow
for effective Yukawa couplings of type      
$Q_LU_R^j H_1$ and $q_L^iD_R^jH_2$ at some level. 
These effective terms could generate the current $u,d,s$-quark masses 
which are all estimated to have values of order or smaller than 
$\Lambda_{QCD}$.

Concerning neutrino masses, since Lepton number is an exact
symmetry Majorana masses are forbidden, there can only be 
Dirac neutrino masses.
The origin of neutrino (Dirac) masses could be quite interesting. 
One expects them to be 
 much more suppressed since neutrinos 
 do not couple directly to strong interaction
effects (which are the source of $U(1)_b$ symmetry breaking). 
In particular, there are in general dimension 6 operators of the
form  $\alpha '(L N_R)( Q_L U_R)^*$. These come from the exchange of
massive string states and are consistent with all
gauge symmetries. Plugging the u-quark  chiral condensate,
  neutrino masses of order
\beq
m_{\nu } \ \propto   \frac{<u_Ru_L>}{M_s^2}
\label{neutrinomass}
\eeq
are obtained
\footnote{The presence of Dirac neutrino masses of this order of magnitude
from this mechanism looks like a general property of low string scale 
models.}
. For $<u_Ru_L>\propto $ $(200\ MeV)^3$ and $M_s\propto 1-10 TeV$
one gets neutrino masses of order $0.1-10 eV$'s, consistent with oscillation 
experiments.  The smallness of neutrino masses 
would be thus related to the existence of a PQ-like symmetry ($U(1)_b$),
which is broken by chiral symmetry breaking.
Notice that the dimension 6 operators may have different coefficients
for different neutrino generations so there will be in general non-trivial
generation structure.

\item
{\it Double Higgs system}

In the case in which both type of Higgs fields $H_i$ and $h_i$
coexist, all quarks and leptons have in general Yukawa 
couplings from the start. The observed hierarchy of fermion masses
would be a consequence of the different values of the Higgs fields 
and hierarchical values for Yukawa couplings. 
In particular, if the vev of the higgs $h_i$ turn out to be small, 
the fermion mass structure would be quite analogous to the previous
case. This could be the case if the Higgs parameters are such 
that the $h_i$ Higgsess were very massive.

\end{itemize}

To reproduce the observed fermion spectrum it is not enough 
with the different mass scales given by the Higgs vevs. Thus for 
example, in the charged lepton sector all masses are proportional to the
vev $<H_2>$ and the hierarchy of lepton masses has to arise from a
hierarchy of Yukawa couplings.
Indeed, in models with intersecting branes it is quite natural 
the appearance of  hierarchical Yukawa couplings.
As was  explained in \cite{afiru2} for the case of D4-branes,
quarks,  leptons and  Higgs fields   
live in general at different intersections. Yukawa couplings among the  
Higgs $h_i,H_i$ and two fermion states $F_R^j$, $F_L^k$ arise from a string
world-sheet stretching among the three D6-branes which cross at those
intersections. The world-sheet has a triangular shape, with vertices on the
relevant intersections, and sides within the D6-brane world-volumes. 
The size of the Yukawa couplings are  of order
\beq
Y_{ijk}\ =\  \exp{(- A_{ijk})}
\label{yuk}
\eeq
where $A_{ijk}$ is the area (in string units) 
 of the world-sheet connecting the three vertices. Since the areas
involved are typically order one in string units, corrections due to
fluctuations of the world-sheet may be important, but we expect the
qualitative behaviour to be controlled by (\ref{yuk}). This structure
makes very natural the appearance of hierarchies in Yukawa couplings of
different fermions, with a pattern controlled by the size
of the triangles. The size of the triangles depends in turn on the
size of the compact radii in the first and third 
tori, $R_{1,2}^{(1)}$ and $R_{1,2}^{(3)}$ but also on the particular 
shape of the triangle. 
The cycle wrapped by a  D6-brane around the i-th torus is
given by a straight line equation
\beq
X_2^{(i)}\ =\ a^{i} (2\pi R_2^{(i)})\ +
\ {{m_i R_2^{(i)}}\over {n^i R_1^{(i)}}}\
X_1^{(i)} \ ,\  i=1,3
\label{str}
\eeq
Thus the area of each triangle depends not only on the wrapping numbers
$(n_i,m_i)$ but also on the $a^i$'s. Since there are four stacks of branes
(plus their mirrors),
there will be all together 8 independent $a^i$ parameters
(in addition to the radii and the different vevs for the Higgs fields
$h_i,H_i$) in order to reproduce the observed quark and lepton spectrum.
It would be interesting to make a systematic analysis of
the patterns of fermion masses in this class of models. We postpone this
analysis to future work.

Concerning the gauge coupling constants, similarly to the 
D4-brane models discussed in \cite{afiru2} they are controlled by the 
length of the wrapping cycles, i.e.,
\beq
{{4\pi ^2}\over {g_i^2} } \ =\
{{M_s}\over {\lambda_{II}}} \ ||l_i||
\label{coup}
\eeq
where $M_s$ is the string scale, $\lambda_{II}$ is the Type II string
coupling, and $||l_i||$ is the length of the cycle of the i-th set of branes
\beq
||l_i||^2\ =\ 
((n_i^1R_1^{(1)})^2+(m_i^1R_2^{(1)})^2)
((n_i^2R_1^{(2)})^2+(m_i^2R_2^{(2)})^2)
((n_i^3R_1^{(3)})^2+(m_i^3R_2^{(3)})^2) \ .
\label{length}
\eeq
Thus, in the case of the SM configurations described in the 
previous sections we have
\beqa
{\alpha_{QCD}}^{-1}\ & = & \ {{M_s}\over {\pi \lambda_{II}}} ||l_a|| \\
\alpha_2^{-1}  \ &  =  & \    {{M_s}\over {\pi \lambda_{II}}} ||l_b|| \\
\alpha_Y^{-1} \ & = & {(6\alpha_{QCD})}^{-1}+
                     {{M_s}\over {\pi \lambda_{II}}}
\frac{1 }{2}( ||l_c||+||l_d||)
\label{coupsm}
\eeqa
where lengths are measured in string units. 
These are the tree level values at the string scale.
In order to compare with the low-energy data one has to 
consider the effect of the running of couplings in between
the string scale $M_s$ and the weak scale. Notice that even
if the string scale is not far away (e.g., if $M_s\propto 1-few $ TeV)
those loop corrections may be important if some of the massive states 
(gonions, windings or KK states) have masses in between 
the weak scale and the string scale. Thus in order to make a full comparison with
experimental data one has to compute the spectra of those massive states
(which depend on radii and intersection angles as well  as the wrapping
numbers of the model considered).  As in the case of Yukawa couplings,
a detailed analysis of each model is required in order to see if
one can reproduce the experimental values. It seems however that there is 
sufficient freedom to accommodate the observed results for some
classes of models.

\section{Final comments and outlook}

In this article we have presented the first string constructions
having just three standard quark/lepton  generations and 
a gauge group $SU(3)\times SU(2)\times U(1)_Y$ from the start.
We have identified a number of remarkable properties which seem more general 
than the specific D6-brane toroidal examples that we have explicitly built.  
In particular: 1) The number of quark-lepton generations is related to 
the number of colours ; 2) Baryon and Lepton numbers are exact 
(gauged) symmetries in
perturbation theory
\footnote{Notice this implies that cosmological baryogenesis
can only happen at the non-perturbative level,
as in weak-scale baryogenesis scenarios.};
 3) There are three generations of right-handed neutrinos
but no Majorana neutrino masses are allowed. 4) There is a gauged 
(generation dependent) Peccei-Quinn-like symmetry.
All these properties depend only on the general structure of 
$U(1)$ anomaly cancellation in a theory of branes with intersection numbers
given by eq.(\ref{intersec2}), yielding just the SM spectrum.
This structure of gauged $U(1)$ symmetries could be relevant 
independently of what the value of the string scale is assumed to be
\footnote{In particular, it is conceivable the existence e.g. of
N=1 supersymmetric models with a string scale of order of the
grand unification mass and with the Baryon, Lepton and 
PQ symmetries gauged in this manner.}.

All these properties are quite interesting. The first offers us a
simple answer to  the famous question "who ordered  the muon".
Anomaly (RR-tadpole) cancellations require more than one generation,
a single standard quark/lepton generation would necessarily have
anomalies in the present context. The second property explains another
remarkable property of the SM, proton stability. With the fermion fields
of the SM one can form dim=6 operators giving rise to proton decay.
The usual explanation for why those operators are so much suppressed 
is to postpone the scale of fundamental (baryon number violating)
physics beyond a scale of order $10^{16}$ GeV. In the present context
there is no need to postpone the scale of fundamental physics to such
high values, the proton would be stable anyhow. This is particularly
important in schemes in which the scale of string theory is 
assumed to be low (1-few TeV) in which up to now there was no 
convincing explanation for the absence of fast proton decay.
The exact conservation of lepton number also gives us important 
information. There cannot exist Majorana neutrino masses 
and hence, processes like  $\nu$-less double beta-decay should
be absent. Neutrino masses, whose existence is supported by
solar and atmospheric neutrino experiments, should be of
Dirac type. Their smallness should not come from a traditional 
see-saw mechanism, given the absence of Majorana masses. 
In the specific toroidal models discussed in the text we give a
possible explanation for their smallness. Due to the presence of
the PQ symmetry in this class of models (which is broken by 
the QCD chiral condensates), a natural scale of order 
$m_{\nu}\propto (\Lambda _{QCD})^3/M_s^2$ appears, which 
is of the correct order of magnitude to be consistent
with the atmospheric neutrino data if the string scale 
is of order 1-few TeV.

The specific examples of SM brane configurations that we construct 
consist on D6-branes wrapping on a 6-torus. We have classified all 
such models yielding the SM spectrum. The analysis of $U(1)$ 
anomaly cancellation is crucial in order to really obtain 
the SM structure in the massless spectrum. We have also shown that
the configurations have no tachyons for wide ranges of the geometric moduli
and hence are stable at this level. 

For certain values of the geometric
moduli one can have extra light fields with the quantum numbers of 
standard Weinberg-Salam doublets which can give rise to 
electroweak symmetry breaking. This implies that the string scale
in this toroidal models should not be  far away from the 
electroweak scale, since the models are non-supersymmetric and
the choice of geometric moduli yielding light Higgs fields would 
become a fine-tuning. As noted in ref.\cite{bgkl} , the usual 
procedure for lowering the string scale down to 1-10 TeV while
maintaining the four-dimensional Plank mass at its experimental
value cannot be applied directly to these D6-brane toroidal models.
This is because if some of the compact radii $R_{1,2}^i, i=1,2,3$
are made large some  charged  KK modes living on the branes would become
very light. The point is that there are not torus directions simultaneously
transverse to all D6-branes. In ref.\cite{afiru,berruga} it was
proposed a way in which one can have a low string scale compatible
with the four-dimensional large Planck mass. The idea is that the
6-torus could be small while being connected to some very large 
volume manifold. For example, one can consider a region of the 
6-torus away from the D6-branes, cut a ball and gluing a throat 
connecting it to a large volume manifold.  In this way one would 
obtain a low string scale model without affecting directly 
the brane structure discussed in the previous sections.
 In the intersecting 
D5 and D4 brane models discussed in refs.\cite{afiru} the 
standard approach for lowering the string scale
with large transverse dimensions can be on the other hand implemented. 
It would be interesting to search for SM configurations in these other
classes of models. Alternatively it could be that 
the apparent large value of the four dimensional Planck mass 
could be associated to the localization of gravity on the branes,
along the lines of \cite{rs}.  This localization could take place at brane 
intersections \cite{kr}.

The D6-brane configurations which we have described are free of
tachyons and RR tadpoles. However the constructions are non-supersymmetric 
and there will be in general NS tadpoles. Thus  the full stability of 
the configurations is an open question. We believe
however that most of our conclusions in the present work are 
a consequence of the chirality of the models and RR-tadpole cancellations
and a final stable configuration should maintain the general 
structure of the models.

The D6-brane toroidal models have also a number of additional
properties of phenomenological interest.
 The light Higgs multiplets are analogous to those
appearing in the MSSM and their number is controlled by the 
integer parameters of the models. We find  
  families of solutions  leading to
the minimal set of Higgs fields or to a double set of Higgs fields,
which would be required if we want all quarks and leptons to have Yukawa
couplings from the start. The structure of the mass terms in the Higgs scalar
potential is quite analogous also to that of the MSSM, although now
the mass parameters have an attractive geometrical interpretation 
in terms of the compactification radii and intersection angles of the
models. Quark and lepton masses depend both on the vevs of the different Higgs
fields and on the properties  of the Yukawa couplings. We showed how the
latter can have a hierarchical structure in a natural way, due to their
exponential dependence on the area of the world-sheet stretching 
among the Higgs fields and the given fermions.
Finally, the gauge couplings are not unified at the string scale in the present
scheme. The size of the couplings are rather inversely proportional to the volume 
each of the branes are wrapping. It would be interesting to see whether
one can find an specific D6-brane model in which we can simultaneously describe
all the observed data on gauge coupling constants and fermion masses,
while obeying experimental limits on extra heavy particles. In particular
the D6-brane toroidal models have extra massive fields on the intersections
(some of them looking like squarks and sleptons) as well as 
KK and winding excitations. Some of these fields may be in the range
in between the weak and the string scale, depending on the compact radii
and intersection angles. 
 A phenomenological study of these extra fields 
would be of interest.

In summary, we have obtained the first string constructions
giving rise {\it just} to three quark/lepton generations   
of the $SU(3)\times SU(2)\times U(1)_Y$ group.
Beyond the particular (quite appealing) features of
the constructed models, we believe the symmetries
of the construction shed  light on  relevant features
 of the standard model like
generation replication, proton stability, lepton number
conservation and other general properties.

\bigskip

\bigskip

\bigskip

\centerline{\bf Acknowledgements}
We are grateful to G. Aldazabal, F.Quevedo and  specially to A.~Uranga for very useful comments and discussions.
The research of R.R. and F.M. was  supported by
 the Ministerio de Educacion, Cultura y Deporte (Spain) through FPU grants.
This work is partially supported by CICYT (Spain) and the 
European Commission (RTN contract HPRN-CT-2000-00148).

\newpage

\section{Appendix I}

\begin{figure}
\centering
\epsfxsize=2in
\hspace*{0in}\vspace*{.2in}
\epsffile{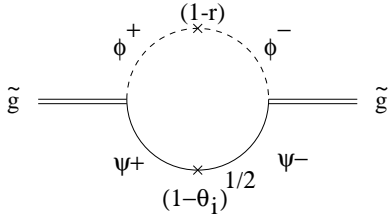}
\caption{\small One loop contribution to gaugino masses.}
\label{oneloopmass2}
\end{figure}

As we mentioned in section 6, all $\cn=4$ massless fields
(with the exception of the gauge group) in the world-volume
of D6-branes become massive in loops. In this appendix 
we show what kind of quantum corrections give masses to
these states from the point of view of the effective Lagrangian.
It is clear that in order for these states to become massive
the loop diagrams have to involve fields living at the intersecting
branes, in which supersymmetry is explicitly broken.
As an example, let us consider the loop corrections giving 
masses to the gauginos. Consider the gauginos in a brane $a$
which intersect other branes labelled by $b$. The effective Lagrangian
diagram contributing to the $a$-branes gaugino masses
is shown in fig.(\ref{oneloopmass2}).

 In order to break chirality only massive fermions
at  the intersections contribute in the loop. Such 
massive fermions exist as we discussed in section 6. 
We will work out for simplicity the case in which 
we are close to one of the $\cn=1$ walls in fig.(\ref{tetra})
where one has an approximate $\cn=1$ supersymmetry unbroken 
at that intersection. For example, consider we were in the vicinity
of $r=1/2(\vartheta_1+\vartheta_2+\vartheta_3)=1$
 Then 
there is a scalar with mass $(1-r)$ which is almost massless,
a $\cn=1$ partner of the chiral fermion at the intersection.
In addition there are three Dirac fermions  (
two Weyl spinors of opposite charges $\Psi^+, \Psi^-$)
with mass$^2$ 
given by $(1-\vartheta_i)$, three complex scalars $T_i$ with masses
$(1-\vartheta_i)-(1-r)$ and other three $T_i'$ with masses
$(1-\vartheta_i)+(1-r)$. This spectrum corresponds to
two $N=1$ chiral supermultiplets 
$\Phi_i^+=(\phi_i^+,\Psi^+)$ and $\Phi_i^-=(\phi_i^-,\Psi^-)$
with $\phi_i^+=1/2(T_i+T_i')$ and $\phi_i^-=1/2(T_i-T_i')$.
The masses of this system in the vicinity of  the $\cn=1$ 
wall may be described by a superspace action 
\beq
\sum_{i=1}^3\ \int d\theta^2\ (m_i\Phi_i^+\Phi_i^-
\ +\ \eta \Phi_i^+\Phi_i^-)
\label{masilles}
\eeq
where $m_i^2=(1-\vartheta_i)$ and $\eta =\theta^2(1-r)$ acts
as a $\cn=1$ SUSY-breaking spurion. It is clear from this structure
that gaugino masses appearing at one loop will be proportional
to the SUSY-breaking parameter $(1-r)$. Indeed, the 
graph in fig.(\ref{oneloopmass2}) contributes to the gaugino masses
(in the limit in which $(1-r)$ is much smaller than $(1-\vartheta_i)$)
\beq
M_a\ =\ \frac{g_a^2}{(4\pi )^2} \frac{(1-r)}{\sqrt{(1-\vartheta_i)}} M_s
\ .
\label{gluino}
\eeq
Notice that this would be just the contribution of one of the
intersections. To get the total contribution one would have to sum
over all intersections. In addition, this is just the contribution of
the lightest set of fermionic and bosonic "gonions". In general 
there is a tower of such massive fields all contributing to this
gaugino masses. Taking into account this, the typical size of
these gaugino masses will be of order the string scale.
Similar loop contributions exist for the other three 
adjoint fermions of the initial massless $\cn=4$ multiplet as well
as for the adjoint scalars. Notice that although, in order to
illustrate the loop corrections we have worked close to a
$\cn=1$ wall in fig.(\ref{tetra}), the general argument remains true 
even if we work in a more generic point.

\newpage

\section{Appendix II}

In this appendix we show the general conditions 
which have to be satisfied in order to
get a SM configuration without tachyonic scalars. 
As already stated in section 5, 
these conditions can be expressed in terms of the 
six angles defined in (\ref{angulos}).
If performed a general analysis, one finds the following 14 conditions:

\beqa
-\theta _1 \ +\ \theta_2 \ +\ {\tilde {\theta_3}} 
\ -\ \theta_3 \ \geq \ 0 \nonumber \\
\theta _1 \ -\ \theta_2 \ +\ {\tilde {\theta_3}} 
\ -\ \theta_3 \ \geq \ 0 \nonumber \\
{\tilde {\theta_1 }}\ -\ {\tilde {\theta_2 }} \ +\ \frac{1}{2}
\ -\  {\tilde {\theta_3}} \ \geq \ 0 \nonumber \\
{\tilde {\theta_1 }}\ +\ {\tilde {\theta_2 }} \ -\ \frac{1}{2}
\ -\  {\tilde {\theta_3}} \ \geq \ 0 \nonumber \\
\frac{3}{2}\ -\ {\tilde {\theta_1 }}\ -\ {\tilde {\theta_2 }} \ 
\ -\  {\tilde {\theta_3}} \ \geq \ 0 \nonumber\\
\nonumber\\
\theta _1 \ +\ {\tilde {\theta_2 }} \ -\ 2{\tilde {\theta_3}} \ 
\geq \ 0  \ &  (if\  \ |n_c^1| < 3\rho n_b^1) \nonumber \\
2\ -\ \theta _1 \ -\ {\tilde {\theta_2 }} \ -\ 2{\tilde {\theta_3}} \ 
\geq \ 0  \ &  (if\  \ |n_c^1| > 3\rho n_b^1) \nonumber\\
\nonumber\\
\left.\begin{array}{r}
\theta _1 \ -\ {\tilde {\theta_2 }} \ +\ 2{\tilde {\theta_3}}  \             
\geq \ 0  \\
{\tilde {\theta_1 }} \ +\ \theta_2 \ -\ \frac{1}{2} \ -\ \theta_3  \             
\geq \ 0 
\end{array}\right\} &  (if\  \ n_a^2 < \rho n_d^2) \nonumber \\
\left.\begin{array}{r}
-\theta _1 \ +\ {\tilde {\theta_2 }} \ +\ 2{\tilde {\theta_3}}  \             
\geq \ 0  \\
\frac{3}{2}\ -\ {\tilde {\theta_1 }} \ -\ \theta_2 \ -\ \theta_3  \             
\geq \ 0 
\end{array}\right\} &  (if\  \ n_a^2 > \rho n_d^2) \nonumber\\
\nonumber\\
 \theta _1 \ +\ \theta_2 \ -\ {\tilde {\theta_3}} \ -\ \theta_3 \  \geq \ 0
& \ (if\ \ |n_c^1| < 3\rho n_b^1 \ \ and \ \ n_a^2 < \rho n_d^2) \nonumber \\ 
\left.\begin{array}{r}
{\tilde {\theta_1 }} \ -\ \theta_2 \ +\ \frac{1}{2} \ -\ \theta_3  \             
\geq \ 0 \\
2 \ -\ \theta _1 \ -\ \theta_2 \ -\ {\tilde {\theta_3}} \ -\ \theta_3  \             
\geq \ 0  \\
\end{array}\right\} &  
\ (if\ \ |n_c^1| > 3\rho n_b^1 \ \ and \ \ n_a^2 > \rho n_d^2)  \
\label{condangulos}
\eeqa
where, if the conditions indicated in brackets are not verified, 
the corresponding constraint is absent. Since some of these conditions
are incompatible, we see that we can have at most ten of them. However,
in most cases many  of the conditions become trivial. If, for instance
we consider models with positive $n_a^2,n_b^1,n_d^2$, then we have that
$\theta_i, \tilde {\theta_i} \leq 1/2$ and $t_4$ scalars are trivially 
massive. In this case our conditions become:
\beqa
-\theta _1 \ +\ \theta_2 \ +\ {\tilde {\theta_3}} 
\ -\ \theta_3 \ \geq \ 0 \nonumber \\
\theta _1 \ -\ \theta_2 \ +\ {\tilde {\theta_3}} 
\ -\ \theta_3 \ \geq \ 0 \nonumber \\
{\tilde {\theta_1 }}\ +\ {\tilde {\theta_2 }} \ -\ \frac{1}{2}
\ -\  {\tilde {\theta_3}} \ \geq \ 0 \nonumber \\
\nonumber\\
\theta _1 \ +\ {\tilde {\theta_2 }} \ -\ 2{\tilde {\theta_3}} \ 
\geq \ 0  \ &  (if\  \ |n_c^1| < 3\rho n_b^1) \nonumber \\
\nonumber\\
\left.\begin{array}{r}
\theta _1 \ -\ {\tilde {\theta_2 }} \ +\ 2{\tilde {\theta_3}}  \             
\geq \ 0  \\
{\tilde {\theta_1 }} \ +\ \theta_2 \ -\ \frac{1}{2} \ -\ \theta_3  \             
\geq \ 0 
\end{array}\right\} &  (if\  \ n_a^2 < \rho n_d^2) \nonumber \\
-\theta _1 \ +\ {\tilde {\theta_2 }} \ +\ 2{\tilde {\theta_3}}  \             
\geq \ 0  \ &  (if\  \ n_a^2 > \rho n_d^2) \nonumber\\
\nonumber\\
 \theta _1 \ +\ \theta_2 \ -\ {\tilde {\theta_3}} \ -\ \theta_3 \  \geq \ 0
& \ (if\ \ |n_c^1| < 3\rho n_b^1 \ \ and \ \ n_a^2 < \rho n_d^2) \ 
\label{condangulos2}
\eeqa
where again bracketed conditions imply the existence 
or not of the constraint. Notice 
that these conditions are expressed only in terms 
of the four integer parameters of 
our models.


\newpage

\begin{thebibliography}{99}



\bibitem{phen}
For reviews on string phenomenology with reference to the original
literature
see e.g.:\\
F. Quevedo, hep-ph/9707434; hep-th/9603074 ;\\
K. Dienes, hep-ph/0004129; hep-th/9602045 ;\\
J.D. Lykken, hep-ph/9903026; hep-th/9607144 ;\\
M. Dine, hep-th/0003175;\\
G. Aldazabal, hep-th/9507162 ;\\
L.E. Ib\'a\~nez, hep-ph/9911499;hep-ph/9804238;hep-th/9505098;\\
Z. Kakushadze and S.-H.H. Tye, hep-th/9512155;\\
I. Antoniadis, hep-th/0102202;\\
E. Dudas, hep-ph/0006190.
%
%
\bibitem{bgkl} 
Ralph Blumenhagen, Lars Goerlich, Boris Kors, Dieter Lust,
Noncommutative Compactifications of Type I Strings on Tori
with Magnetic Background Flux, JHEP 0010 (2000) 006, hep-th/0007024 \\
Magnetic Flux in Toroidal Type I Compactification, hep-th/0010198.
%
\bibitem{afiru}
G.~Aldazabal, S.~Franco, L.E.~Iba\~{n}ez, R.~Rabadan, A.M.~Uranga,
D=4 Chiral String Compactifications from Intersecting Branes, hep-th/0011073.%
%
\bibitem{afiru2}
G.~Aldazabal, S.~Franco, L.E.~Iba\~{n}ez, R.~Rabadan, A.M.~Uranga,
Intersecting brane worlds, hep-ph/0011132.
%
\bibitem{bkl}
Ralph Blumenhagen, Boris Kors, Dieter Lust,
Type I Strings with F- and B-Flux,
 JHEP 0102 (2001) 030, hep-th/0012156.
%
%
\bibitem{bachas}
C.Bachas, A way to break supersymmetry, hep-th/9503030.
%
%
\bibitem{bfield}
M. Bianchi, G. Pradisi and A. Sagnotti,
Toroidal Compactification and Symmetry Breaking in Open-String Theories,
\NPB{376}{92}{365} ;\\
M. Bianchi, A Note on Toroidal Compactifications of the Type I 
Superstring and Other Superstring Vacuum Configurations with 16 Supercharges,
\NPB{528}{98}{73}, hep-th/9711201;\\
E. Witten, Toroidal Compactification Without Vector Structure,
JHEP  9802(1998)006, hep-th/9712028;\\
C. Angelantonj, Comments on Open String Orbifolds
with a Non-Vanishing $B_{ab}$, hep-th/9908064;\\
Z. Kakushadze, Geometry of Orientifolds with NS-NS B-flux,
Int.J.Mod.Phys.  A15(2000)3113,hep-th/0001212;\\
C. Angelantonj and A. Sagnotti, Type I Vacua and Brane Transmutation,
hep-th/0010279.
%
\bibitem{flux}
C. Angelantonj, I. Antoniadis, E. Dudas, A. Sagnotti,
Type-I strings on magnetised orbifolds and brane
transmutation, Phys.Lett. B489 (2000) 223-232, hep-th/0007090. 
%
\bibitem{gs}
M. Green and J.H. Schwarz, \PLB{149}{84}{117}.
%
\bibitem{lykken}
J.~D.~Lykken, Phys. Rev. D54 (1996) 3693, hep-th/9603133.
%
\bibitem{aadd}
N. Arkani-Hamed, S. Dimopoulos and G. Dvali, Phys. Lett. B429 (1998) 263,
hep-ph/9803315;\\
I. Antoniadis, N. Arkani-Hamed, S. Dimopoulos, G. Dvali
\PLB{436}{99}{257}, hep-ph/9804398.
%
\bibitem{otherbw}
K. Dienes, E. Dudas and T. Gherghetta, Phys. Lett. B436 (1998) 55,
hep-ph/9803466;\\
R. Sundrum, Phys.Rev. D59 (1999) 085009, hep-ph/9805471; Phys. Rev. D59
(1999) 085010, hep-ph/9807348;\\
G.~ Shiu, S.H. Tye, \PRD{58}{98}{106007}, hep-th/9805157;\\
Z. ~Kakushadze, \PLB {434}{98}{269}, hep-th/9804110; \PRD{58}{98}101901,
hep-th/9806044;\\
C. Bachas, JHEP 9811 (1998) 023, hep-ph/9807415;\\
Z. Kakushadze, S.H. Tye, Nucl.Phys. B548 (1999) 180, hep-th/9809147;\\
K.~Benakli, Phys. Rev. D60 (1999) 104002, hep-ph/9809582; \\
C.P. Burgess, L.E. Ib\'a\~nez, F. Quevedo, \PLB{447}{99}{257},
hep-ph/9810535.\\
L.E. Ib\'a\~nez, C. Mu\~noz, S. Rigolin, \NPB{553}{99}{43},
hep-ph/9812397.\\
A. Delgado, A. Pomarol and M. Quiros, Phys. Rev. D60 (1999) 095008,
hep-ph/9812489;\\
L.E. Ib\'a\~nez and F. Quevedo, hep-ph/9908305;\\
E.~Accomando, I.~Antoniadis, K.~Benakli,
Nucl. Phys. B579 (2000) 3, hep-ph/9912287;\\
D. Ghilencea and G.G. Ross, Phys.Lett.B480 (2000) 355, hep-ph/0001143;\\
I. Antoniadis, E. Kiritsis and T. Tomaras, Phys.Lett.
B486 (2000) 186, hep-ph/0004214;\\
S. Abel, B. Allanach, F. Quevedo, L.E. Ib\'a\~nez and M. Klein,
hep-ph/0005260.
%
\bibitem{pq}
R. Peccei and H. Quinn, \PRL{38}{77}{1440};\\
S. Weinberg, \PRL{40}{78}{223};\\
F. Wilczek, \PRL{40}{78}{278}.
%
\bibitem{evenmore}
G.~Aldazabal, L.~E.~Ib\'a\~nez, F.~Quevedo, JHEP 0001 (2000) 031,
hep-th/9909172; JHEP02 (2000) 015, hep-ph/0001083.\\
M.~Cvetic, A.~M.~Uranga, J.~Wang,
hep-th/0010091.
%
\bibitem{aiqu}
G.~Aldazabal, L.~E.~Ib\'a\~nez, F.~Quevedo, A.~M.~Uranga,
D-branes at singularities: A Bottom up approach to the string embedding
 of the standard model,
JHEP 0008:002,2000. [hep-th/0005067]
%
\bibitem{bjl}
D. Berenstein, V. Jejjala and R.G. Leigh,
The Standard Model on a D-brane,
hep-ph/0105042.
%
\bibitem{orientold}
A.~Sagnotti, in Cargese 87, {\it Strings  on Orbifolds}, ed. G.
Mack et al. (Pergamon Press, 1988) p. 521; P.~Horava, \NPB{327} {89}
{461}; J.~Dai, R.~Leigh and J.~Polchinski, Mod.Phys.Lett. A4 (1989) 2073;
G.~Pradisi and A.~Sagnotti, \PLB{216} {89} {59} ; M.~Bianchi and
A.~Sagnotti, \PLB{247} {90} {517} ; Nucl. Phys. B361 (1991) 519.
%
\bibitem{orientnew}
E.~Gimon and J.~Polchinski, Phys.Rev. D54 (1996) 1667, hep-th/9601038;
E.~Gimon and C.~Johnson, \NPB{477}{96}{715}, hep-th/9604129;
A.~Dabholkar and J.~Park, \NPB{477}{96}{701}, hep-th/9604178.
%
\bibitem{dsw}
M. Dine, N. Seiberg and E. Witten, \NPB{289}{87}{589};
J. Atick, L. Dixon and A. Sen, \NPB{292}{87}{109};
M. Dine, I. Ichinose and N. Seiberg, \NPB{293}{87}{253}.
%
\bibitem{sagnan}
A.~Sagnotti,
A Note on the Green-Schwarz mechanism in open string theories,
Phys. Lett. B294 (1992) 196, hep-th/9210127.
%
\bibitem{iru}
L.~E.~Ib\'a\~nez, R.~Rabad\'an, A.~M.~Uranga,
Anomalous U(1)'s in type I and type IIB D = 4, N=1 string vacua,
Nucl.Phys. B542 (1999) 112-138, hep-th/9808139.
%
%
\bibitem{bdl}
M.~Berkooz, M.~R.~Douglas, R.~G.~Leigh,
Branes intersecting at angles,
Nucl.Phys. B480(1996)265, hep-th/9606139.
%
\bibitem{angles}
M.M. Sheikh-Jabbari,
Classification of Different Branes at Angles,
Phys.Lett. B420 (1998) 279-284, hep-th/9710121. \\
H. Arfaei, M.M. Sheikh-Jabbari,
Different D-brane Interactions,
Phys.Lett. B394 (1997) 288-296, hep-th/9608167 \\
Ralph Blumenhagen, Lars Goerlich, Boris Kors,
Supersymmetric Orientifolds in 6D with D-Branes at Angles,
Nucl.Phys. B569 (2000) 209-228,
hep-th/9908130; \\
Ralph Blumenhagen, Lars Goerlich, Boris Kors,
Supersymmetric 4D Orientifolds of Type IIA with D6-branes at
Angles,
JHEP 0001 (2000) 040,hep-th/9912204; \\
Stefan Forste, Gabriele Honecker, Ralph Schreyer,
Supersymmetric $Z_N \times Z_M$ Orientifolds in 4D with
D-Branes at Angles,
Nucl.Phys. B593 (2001) 127-154, hep-th/0008250. \\
Ion V. Vancea,
Note on Four Dp-Branes at Angles,
JHEP 0104:020,2001,
hep-th/0011251.
%
\bibitem{probes}
A.M.~Uranga,
D-brane probes, RR tadpole cancellation and K-theory charge,
Nucl.Phys. B598 (2001) 225-246.
%
\bibitem{witten}
M. Mihailescu, I.Y. Park, T.A. Tran
D-branes as Solitons of an N=1, D=10 Non-commutative Gauge Theory,
hep-th/0011079 \\
E. ~Witten,
BPS Bound States Of D0-D6 And D0-D8 Systems In A B-Field, hep-th/0012054. 
%
\bibitem{kachru}
Shamit Kachru, John McGreevy,
Supersymmetric Three-cycles and (Super)symmetry Breaking,
Phys.Rev. D61 (2000) 026001.
%
\bibitem{bbh}
Ralph Blumenhagen, Volker Braun, Robert Helling,
Bound States of D(2p)-D0 Systems and Supersymmetric p-Cycles,
hep-th/0012157.
%
\bibitem{torons}
G. 't Hooft, Nucl.Phys. B153 (1979) 141; Comm. Math. Phys. 81 (1981) 257; \\
P. Van Baal, Comm. Math. Phys. 94 (1984) 397; Comm. Math. Phys. 85 (1982) 529;
\\
J. Troost, Constant field strengths on $T^{2n}$, Nucl.Phys. B568 (2000) 180-194; \\
J. Bogaerts, A. Sevrin, J. Troost, W. Troost, S. van der Loo,
D-branes and constant electro-magnetic backgrounds, hep-th/0101018; \\
Z. Guralnik, S. Ramgoolam, From 0-Branes to Torons, Nucl.Phys. B521 (1998)
129-138.
%
\bibitem{D7}
Edward Witten,
D-Branes And K-Theory,
JHEP 9812 (1998) 019, hep-th/9810188.\\
R. Rabadan, A. M. Uranga
Type IIB Orientifolds without Untwisted Tadpoles, and non-BPS D-branes, 
JHEP 0101 (2001) 029. \\
O. Loaiza-Brito, A.M. Uranga,
The fate of the type I non-BPS D7-brane,
hep-th/0104173. 
%
\bibitem{ir}
L.E. Ib\'a\~nez and G.G. Ross, Phys.Lett.B110(1982)215.
%
\bibitem{berruga}
A. Uranga, unpublished (2000).
%
\bibitem{rs}
L. Randall and R. Sundrum, 
An Alternative to Compactification, Phys.Rev.Lett.83(1999)4690, 
hep-th/9906064.
%
\bibitem{kr}
A. Karch and L. Randall,
Localized Gravity in String Theory,
hep-th/0105108.
%




 
 
\end{thebibliography}
\end{document}